\documentclass[11pt,a4paper]{article}
\pdfoutput=1
\usepackage{graphicx} 
\usepackage[utf8]{inputenc}
\usepackage{a4wide}
\usepackage{amsmath}
\usepackage{amssymb}
\usepackage{amsfonts}
\usepackage{cite}
\usepackage{color}
\usepackage{adjustbox}
\usepackage{multirow}
\usepackage{pdflscape}
\usepackage{textcomp}
\usepackage{gensymb}
\usepackage{diagbox}
\usepackage{pifont}
\usepackage{bigstrut}
\usepackage[small,bf]{caption}
\usepackage{tabulary}
\usepackage{booktabs}
\usepackage{stackengine}
\usepackage{mathtools}
\usepackage{enumerate}
\usepackage{makecell}
\usepackage{listings}
\usepackage{bbm}
\usepackage[unicode]{hyperref}

\definecolor{greenLinks}{rgb}{0, 0.6, 0} 
\definecolor{blueLinks}{rgb}{0, 0, 0.6}
\definecolor{redLinks}{rgb}{0.6, 0, 0}
\definecolor{eprintLinks}{rgb}{0.4, 0.4, 0.4}
\definecolor{journalLinks}{rgb}{0.6, 0, 0}
\newcommand{\be}{\begin{equation}}
\newcommand{\ee}{\end{equation}}
\newcommand{\nua}[1]{\ensuremath{\rlap{\kern-2.5pt\ensuremath{\overset{\scriptscriptstyle(-)}{\phantom{\nu}}}}{\ensuremath{{\nu}_{#1}}}}}
\def\D{\Delta}

\def\f{\Phi}

\DeclareMathOperator{\Tr}{Tr}
\DeclareMathOperator{\diag}{diag}
\DeclareMathOperator{\im}{Im}
\DeclareMathOperator{\re}{Re}
\allowdisplaybreaks
\numberwithin{equation}{section}
\makeatletter
\g@addto@macro\bfseries{\boldmath}
\makeatother
\begin{document}
%%%%%%%%%%%%%%%%%%%%%%%

%%%%%%%%%%%%%%%%%%%%%%%
\begin{titlepage}

\vspace*{8mm}

\begin{center}
{\bf\LARGE {Prospects of light sterile neutrino searches 
 in \\[2mm] long-baseline neutrino oscillations}}\\[8mm]
Yakefu Reyimuaji $^{\,a,}$\footnote{E-mail: \texttt{yakefu@mail.itp.ac.cn}}, 
Chun Liu $^{\,a,b,}$\footnote{E-mail: \texttt{liuc@mail.itp.ac.cn}}\\ 

\vspace{8mm}
$^{a}$\,{\it CAS Key Laboratory of Theoretical Physics, Institute of Theoretical Physics, Chinese Academy of Sciences, Beijing 100190, China} \\
\vspace{2mm}
$^{b}$\,{\it School of Physical Sciences, University of Chinese Academy of Sciences, Beijing 100049, China} \\
\vspace{2mm}
 
\end{center}
\vspace{8mm}

\begin{abstract}
\noindent
The neutrino oscillation probabilities in vacuum and matter are discussed, considering the framework of three active and one light sterile neutrinos. We study in detail the rephasing invariants and CP asymmetry observables, and investigate the four-neutrino oscillations in long-baseline neutrino experiments, such as DUNE, NO$\nu$A and T2HK. Our results show that the matter effect enhances quite a significantly the oscillation probabilities of electron neutrino and electron antineutrino appearance channels within a certain energy range, while no considerable change arises in the CP asymmetry analysis due to the matter effect. Moreover, separation between the results with and without the sterile neutrino is not so significant and that is also affected by CP-violating phases. Comparing the results for these three experiments, all of them have similar features, nevertheless, sizes and separations of the oscillation probabilities in DUNE are bit larger. 

\end{abstract}

\end{titlepage}
\setcounter{footnote}{0}
%%%%%%%%%%%%%%%%%%%%%%%

%%%%%%%%%%%%%%%%%%%%%%%
\section{Introduction}
\label{sec:intro}
%%%%%%%%%%%%%%%%%%%%%%%

Known for their invisible and illusive behaviors, originated from very weak interaction strength and being the lightest and electrically neutral fermion observed so far, neutrinos are one of the most appealing elementary particles in the present-day physics. Along with having made quite a number of  progresses in understanding the properties of three types neutrinos, many things have ceased to become mystery anymore. But still there have been some puzzles and bottlenecks looming ahead and several questions have yet to be answered. In particular, there are several compelling anomalies that cannot be explained within the standard paradigm of three-neutrino mixing: 1) unexpected excess of $\bar{\nu}_e$ events in $\bar{\nu}_\mu \to \bar{\nu}_e$ transition observed by the LSND experiment~\cite{Aguilar:2001ty} as well as an excess discovered in both $\nu_e$  and $\bar{\nu}_e$ channels in the MiniBooNE experiment~\cite{Aguilar-Arevalo:2018gpe}, which are so called short-baseline anomalies; 2) observed rate deficit of $\bar{\nu}_e$ in several reactor experiments compared to theoretical expectation~\cite{Mention:2011rk,Mueller:2011nm,Huber:2011wv}; 3) a rate deficit in the disappearance channel of $\nu_e$ in radiochemical experiments, such as GALLEX~\cite{Hampel:1998xg,Kaether:2010ag} and SAGE~\cite{Abdurashitov:1999zd,Abdurashitov:2009tn}, using gallium as a target in detector, which is referred to as the gallium anomaly. An interesting fact is that explanations of these anomalies hint towards the existence of a fourth neutrino state, which is uncharged under the weak interaction and thus known as a sterile neutrino, having a few percent mixing with the electron neutrino and a mass of around 1eV. It is important to check if such a sterile neutrino exists from long-baseline neutrino oscillation experiments like DUNE~\cite{Acciarri:2016crz,Acciarri:2015uup,Strait:2016mof,Acciarri:2016ooe}, NO$\nu$A~\cite{Ayres:2007tu} and T2HK~\cite{Abe:2015zbg} since main features of these experiments include precision measurements of neutrino oscillation parameters and CP violation. 

There have been great amount of efforts spent on understanding the possible origins of generating the eV-scale mass and on various experimental searches for the effects of the light sterile neutrino. Mechanisms to generate an eV-scale sterile neutrino mass have been provided, such as in Refs.~\cite{Liu:1998qp,Barry:2011wb,Kawai:2019uei}. Based on the simulations of long-baseline experiments, effects of the sterile neutrino on the oscillation probabilities of $\overset{\scriptscriptstyle(-)}{\nu}_e$ appearance and $\nu_\mu$ disappearance channels, the measurement of CP violation, the neutrino mass hierarchy, the octant of $\theta_{23}$, and determination of CP-violating phases have been studied in~\cite{Berryman:2015nua,Gandhi:2015xza,Agarwalla:2016mrc,Agarwalla:2016xxa,Dutta:2016glq,Kelly:2017kch,Choubey:2017cba,Coloma:2017ptb,Choubey:2017ppj,Gupta:2018qsv,Choubey:2018kqq,Betti:2019ouf}. Using the available data, experimental constraints on the sterile neutrino oscillation parameters have been investigated in~\cite{Lindner:2015iaa,Abe:2019fyx}. On the other hand, in Ref.~\cite{Kamo:2002sj} authors have derived analytically the four-neutrino oscillation probabilities in matter by assuming all the CP-violating phases vanish. The same has been studied in~\cite{Li:2018ezt} with different approach and taking into account also the CP violation. A perturbative calculation method has been extended to 3+1 neutrino oscillation scenario \cite{Parke:2019jyu} provided that the matter potential is smaller or comparable to the atmospheric mass-squared difference. Current status of the sterile neutrinos is reviewed in~\cite{Abazajian:2012ys,Boser:2019rta,Diaz:2019fwt}. These consecutive works in this direction have put the light sterile neutrino at the focus of neutrino physics research.

In this work, we analyze systematically the 3+1 neutrino scenario in long-baseline oscillations. The full analytic formulae for four-neutrino oscillation probabilities in matter will be derived, and numerical results will be given in detail. Although Ref.~\cite{Li:2018ezt} provided analytic expressions, our results are different from them. The numerical results given in Refs.~\cite{Berryman:2015nua,Gandhi:2015xza,Agarwalla:2016mrc,Agarwalla:2016xxa,Dutta:2016glq,Kelly:2017kch,Choubey:2017cba,Coloma:2017ptb,Choubey:2017ppj,Gupta:2018qsv,Choubey:2018kqq,Betti:2019ouf} can be directly obtained from our
analytic expressions. This is also a check of the correctness of our results. Furthermore, it is desirable to explore all possible oscillation channels, based on the exact analytic results, including neutrinos and antineutrinos, to show the effects of the sterile neutrino in the several long-baseline neutrino oscillations. And it is also worth carrying out a systematic theoretical study of CP asymmetry in four-flavor oscillations in a more general and accurate way, since that is useful to make an expectation about the results from the future long-baseline experiments. We also provide extensive discussions on the Jarlskog invariants and CP-violating observables in the case of three active plus one sterile neutrinos, and apply our results to three long-baseline experiments -- DUNE , NO$\nu$A and T2HK -- to see distinctions of the 3+1 neutrino oscillation and CP asymmetry compared to the standard three-neutrino scenario. Of course, one can use our results to any oscillation channels of three active and one sterile neutrinos, regardless of the values of sterile neutrino mass and mixing, as such they are useful for any long-baseline neutrino oscillation experiments. We also try to put our work in the more general ground and to carry out analysis in the parameterization independent way of mixing matrix. 

The structure of this paper is organized in the following order. In section~\ref{sec:NuOscillation}, we investigate the 3+1 neutrino oscillation, including matter effects to neutrino mass squares and to products of one mixing matrix element with complex conjugate of any other element in the same column. Section~\ref{sec:CPviolation} describes Jarlskog invariants and CP asymmetries with four neutrinos and provides relations between their values in matter and vacuum. Then, in section~\ref{sec:nusrch@DUNE}, there is a phenomenological study about the neutrino oscillations including the eV-scale light sterile neutrino and CP asymmetry observables in the long-baseline experiments such as DUNE, NO$\nu$A and T2HK. Section~\ref{sec:conclusions} is the conclusion of the paper. At the end, there are several appendices titled~\ref{app:sol4polyeq}, \ref{app:sumrld4mixing} and \ref{app:jcpexp} that fill gaps in our derivations and give supplementary information.

%%%%%%%%%%%%%%%%%%%%%%%
\section{Matter effects in 3+1 neutrino scheme}
\label{sec:NuOscillation}
%%%%%%%%%%%%%%%%%%%%%%%

In the three-neutrino paradigm, the flavor mixing is well understood and described by the PMNS matrix, which is a $3 \times 3$ unitary rotation matrix to go from mass eigenstate basis to flavor basis. When it is extended to 3+1 scenario, by adding one sterile neutrino, corresponding mixing matrix has to be introduced. There are many possible ways to parameterize this $4\times 4$ mixing matrix in terms of products of unitary rotations on two-dimensional planes, which are made up of the Euler angles and phases. One convenient way to write its form is the following
\be\label{eq:U4mixmat}
U = U_s U_3,
\ee
 where the matrix $U_s$ encompasses the mixing between the sterile and active neutrinos, while the $U_3$ is an embedding of PMNS matrix into the upper left block:
\begin{align}
 U_s = \, & R_{34}(\theta_{34},0)R_{24}\left(\theta_{24},\delta^{CP}_{24}\right)R_{14}\left(\theta_{14},\delta^{CP}_{14}\right)\, , \notag \\
 U_3 = \, &R_{23}(\theta_{23},0)R_{13}\left(\theta_{13},\delta^{CP}_{13}\right)R_{12}(\theta_{12},0)\, ,
\end{align}
in which $0\le \theta_{ij} \le \pi/2$ and $0\le \delta^{CP}_{ij} < 2\pi$. An $mn$ element of these rotations on two dimensional planes, on the right-hand sides of above expressions, reads
\begin{align}
\left[R_{ij}(\theta_{ij},\delta^{CP}_{ij})\right]_{mn}= & \left(\delta_{im}\delta_{in}+\delta_{jm}\delta_{jn}\right)\cos\theta_{ij}+\left(\delta_{im}\delta_{jn}e^{-i\delta^{CP}_{ij}}-\delta_{in}\delta_{jm}e^{i\delta^{CP}_{ij}}\right)\sin\theta_{ij}\notag \\
& +\sum_{k\neq i,j}^{4}\delta_{km}\delta_{kn}\, .
\end{align}
Additionally, the mixing matrix in eq.~\eqref{eq:U4mixmat} carries a diagonal matrix of three phases, $\diag(1,$ $ e^{i\alpha_{21}/2}, e^{i\alpha_{31}/2},e^{i\alpha_{41}/2} )$, in the rightmost position if neutrinos are Majorana particles. But these phases are not relevant for neutrino oscillations.

At the present stage, active neutrino oscillation, a flavor change during the propagation from source to detector, is interpreted as a consequence of non-zero masses and mixing angles. Probability of flavor eigenstate $\nu_\alpha$ at a source converted into flavor eigenstate $\nu_\beta$ after some distance $L$ is given by
\begin{align} \label{eq:nuosprob}
P_{\nu_\alpha \to \nu_\beta}(L,E)= \ &\delta_{\alpha\beta}-4\sum_{i < j}\re \left(U_{\alpha i}U_{\alpha j}^*U_{\beta i}^*U_{\beta j}\right)\sin^2\frac{\Delta m^2_{ji} L}{4E} \notag \\
& +2\sum_{i < j}\im \left(U_{\alpha i}U_{\alpha j}^*U_{\beta i}^*U_{\beta j}\right)\sin\frac{\Delta m^2_{ji} L}{2E}\, ,
\end{align}
where $E$ is neutrino energy and  $\Delta m^2_{ji}\equiv  m^2_j- m^2_i $ . Regarding the oscillation between antineutrinos, CPT invariance of the probability implies that $P_{\nu_\alpha \to \nu_\beta}(L,E)=P_{\bar{\nu}_\beta \to \bar{\nu}_\alpha}(L,E)$. This, in turn, leads to the conclusion: when $\nu_\alpha$ and $\nu_\beta$ in the oscillation probability above are replaced by corresponding antiparticles, the signs of all but last term remain the same. Here, and in what follows, the Greek alphabets $\alpha, \beta, \gamma, \dots$ indicate flavor eigenstates of electron, muon, tauon and sterile neutrino types, while the Latin letters $i, j, k, \dots$ are assigned to mass basis, running from 1 to 4. The expression \eqref{eq:nuosprob} holds not only for any number of neutrino flavors, but also for the case of any kind of environment neutrinos go through, by replacing the mass and mixing parameters accordingly. It is sometimes written in slightly different form by changing the first two terms with the help of an identity 
\be
\delta_{\alpha\beta} = \sum_i \left| U_{\alpha i}\right|^2 \left| U_{\beta i}\right|^2 + 2\sum_{i < j}\re \left[ U_{\alpha i}U_{\alpha j}^*U_{\beta i}^*U_{\beta j}\right] ,
\ee
which we will use when discussing the neutrino oscillation probabilities in matter.

While neutrino is propagating in a medium, a coherent scattering with other particles through charged current (CC) as well as neutral current (NC) weak interactions changes its motion. Thus, effective Hamiltonian is a sum of vacuum and matter contributions:
\begin{align}\label{eq:releffham}
 H_\mathrm{eff} = H_0+V .
\end{align}
The vacuum Hamiltonian $H_0$ and components of the matter potential $V$, generated by a uniform matter density, are 
\begin{align}
H_0 & = \frac{1}{2E} U \diag(m_1^2, m_2^2, m_3^2, m_4^2) U^\dagger, \notag \\
V_{\alpha\beta} & = V_\mathrm{CC} \delta_{\alpha e}\delta_{\beta e} +V_\mathrm{NC}\left(\delta_{\alpha \beta}-\delta_{\alpha s}\delta_{\beta s}\right) \notag\\
&=\sqrt{2}G_F\left[ N_e \delta_{\alpha e}\delta_{\beta e} -\frac{1}{2} N_n \left(\delta_{\alpha \beta}-\delta_{\alpha s}\delta_{\beta s}\right) \right],
\end{align}
respectively. As one can see, the potential for sterile neutrino is vanishing since it does not participate in any interactions, apart from the gravity. The CC potential $V_\mathrm{CC}$ depends only on electron number density, $N_e$, due to the interaction with electrons in matter, while the NC potential $V_\mathrm{NC}$ is a function of neutron number density, $ N_n$, only as there is a cancellation between the potentials of electron and proton in neutral matter. Clearly, this matter potential has a contribution to neutrino evolution equation. But one can see, to consider the Earth's matter effect especially when conducting long-baseline oscillation experiments, that its contribution is not so large. This is because $G_F = 5.37\times 10^{-14} \mathrm{eV} \mathrm{cm}^3/N_A$, in which $N_A$ is Avogadro constant, and the range of mean electron number density from the Earth's mantle to core is $2.2 - 5.4 \mathrm{cm}^{-3} N_A$ \cite{Tanabashi:2018oca}. Neutron number density is approximately equal to the number density of electron.  As a result, the effective potential is as large as $10^{-13}$ eV, it might have significant effect at higher energies as the vacuum Hamiltonian $H_0$ is inversely proportional to the neutrino energy $E$. We will see this clearly when we carry out a phenomenological study in section~\ref{sec:nusrch@DUNE}.

Focusing on the neutrino oscillation in matter, the effective Hamiltonian can be simplified by subtracting the potential $V_\mathrm{NC}$. The reason is that subtracting a multiple of an identity matrix from the Hamiltonian is equivalent to a phase shift of transition amplitude from one neutrino flavor eigenstate to another, under which oscillation probability remains the same. So, without causing any problem, in the following discussions we will use an effective Hamiltonian
\begin{align}
 H_\mathrm{eff} = \frac{1}{2E} U \diag(m_1^2, m_2^2, m_3^2, m_4^2) U^\dagger + \diag\left(V_\mathrm{CC}, 0, 0, -V_\mathrm{NC}\right) \, .
 \label{eq:effham1}
\end{align}

As was stated before, the Hamiltonian $H_0$ governs neutrino evolution in vacuum and the unitary matrix that diagonalizes it enters oscillation probability in eq.~\eqref{eq:nuosprob}. Similarly, one can diagonalize the matter-induced effective Hamiltonian $H_\mathrm{eff}$ by unitary rotation $\tilde{U}$
\be
H_\mathrm{eff} = \frac{1}{2E} \tilde{U} \diag(\tilde{m}_1^2, \tilde{m}_2^2, \tilde{m}_3^2, \tilde{m}_4^2) \tilde{U}^\dagger\, ,
 \label{eq:effham2}
\ee
and interpret $\tilde{m}_i^2$ and  $\tilde{U}$ respectively as effective mass squares and mixing matrix in matter. Being a hermitian matrix, diagonalization of $H_\mathrm{eff}$ is rather straightforward and effective mass squares $\tilde{m}_i^2$, which are the eigenvalues, satisfy quartic polynomial equation
\be
(\tilde{m}_i^2)^4 - c_3 (\tilde{m}_i^2)^3 + c_2 (\tilde{m}_i^2)^2 - c_1 \tilde{m}_i^2  + c_0 =0 ,
\ee
where $c_3, c_2, c_1, c_0$ are respectively the trace, the sum of determinants of $2\times 2$ main diagonal blocks, the sum of determinants of $3\times 3$ main diagonal blocks and the determinant of $2EH_\mathrm{eff}$. Their explicit expressions can be found in appendix~\ref{app:sol4polyeq}.
We find following solutions of this quartic equation 
\begin{align}
\tilde{m}_1^2 & =  \frac{c_3}{4} - \frac{1}{2} \left[ \sqrt{2z} + \sqrt{-2z -2p + \sqrt{\frac{2}{z} } \, q \;}  \; \right] , \notag \\
\tilde{m}_2^2 & =  \frac{c_3}{4} - \frac{1}{2} \left[ \sqrt{2z} - \sqrt{-2z -2p + \sqrt{\frac{2}{z} } \, q \;}  \; \right] , \notag \\
\tilde{m}_3^2 & =  \frac{c_3}{4} + \frac{1}{2} \left[ \sqrt{2z} - \sqrt{-2z -2p - \sqrt{\frac{2}{z} } \, q \;}  \; \right] , \notag \\
\tilde{m}_4^2 & =  \frac{c_3}{4} + \frac{1}{2} \left[ \sqrt{2z} + \sqrt{-2z -2p - \sqrt{\frac{2}{z} } \, q \;}  \; \right] ,
\end{align}
where
\begin{align*}
p & = \frac{8 c_2 -3 c^2_3}{8} , \notag \\
q & = -\frac{ c^3_3 - 4 c_2 c_3 + 8 c_1}{8} \, ,
\end{align*} 
and $z$ is a solution of the cubic equation
\be
z^3 + p z^2 + \frac{1}{4} \left( p^2 -4r\right)z -\frac{1}{8}q^2 = 0 \, ,
\ee
where the parameter $r$ is defined by
\begin{equation*}
 r = \frac{-3 c^4_3 + 256 c_0 - 64 c_1 c_3 + 16 c_2 c^2_3}{256} .
\end{equation*}
An explicit form of $z$ is given in the appendix~\ref{app:sol4polyeq}. Here we list effective mass squares in increasing order, under the default assumption that the active neutrino masses are in the normal ordering. If it is aimed to consider the case of inverted mass ordering, one can replace $\tilde{m}_1^2$ by $\tilde{m}_3^2$, $\tilde{m}_2^2$ by $\tilde{m}_1^2$ and $\tilde{m}_3^2$ by $\tilde{m}_2^2$, without spending much effort.

\begin{figure}[ht]
  \centering
  \includegraphics[width=0.495\textwidth]{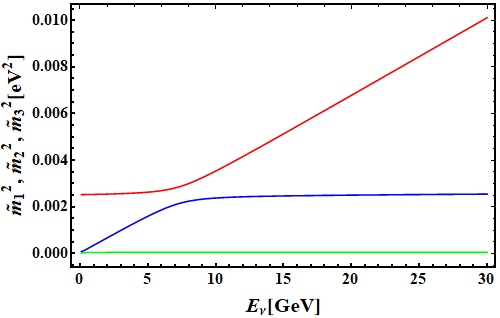}
  \includegraphics[width=0.495\textwidth]{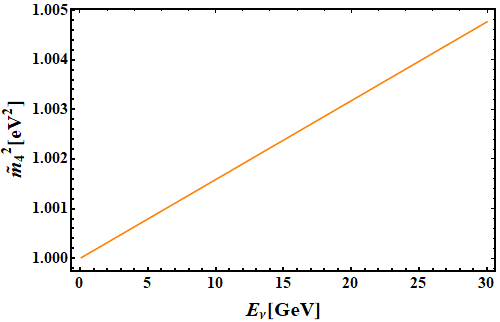}
  \caption{
    Effective mass squares induced by neutrino interaction with matter,  $\tilde{m}^2_1$ in green,  $\tilde{m}^2_2$ in blue,  $\tilde{m}^2_3$ in red and  $\tilde{m}^2_4$ in orange.}
  \label{fig:masssqr}
\end{figure}

Dependences of the effective mass squares on the neutrino energy are depicted in figure~\ref{fig:masssqr}. To make this figure, we simply assume that the lightest neutrino mass is zero, and used the best fit values for active neutrino mass-squared differences $\D m^2_{21} = 7.39 \times 10^{-5} \, \mathrm{eV}^2$, $\left|\D m^2_{3l}\right| = 2.53 \times 10^{-3} \mathrm\, \mathrm{eV}^2$ from \cite{Esteban:2018azc} and $m_4 = 1$ eV, $\sin^2 \theta_{14}\simeq 0.01$ from \cite{Boser:2019rta}. As shown in the left panel, the first effective mass square $\tilde{m}^2_1$ is almost zero and does not change over the neutrino energy, while $\tilde{m}^2_2$ increases up to 8 GeV then starts to saturate with a value around $0.0025$ $\text{eV}^2$. Other two mass squares $\tilde{m}^2_3$ (in the left panel) and $\tilde{m}^2_4$ (in the right panel) increase monotonically. It is worth noticing that these plots can also be interpreted in a different way. Although they are plotted as a function of $E$, in practice the combinations $E V_\mathrm{CC}$ and $E V_\mathrm{NC}$ always enter to the expressions of the effective mass squares. Here we showed the change of $\tilde{m}^2_i$ as a function of energy by fixing the matter potentials, namely assuming constant matter density. On the other hand, one can also interpret the behavior of these effective mass squares due to the fixing of neutrino energy and gradually increase of the matter density (or matter potential as a whole).
%
%%%%%%%%%%%%%%%%
%%%%%%%%%%%%%%%%
%

\begin{figure}%[h]
  \centering
  \includegraphics[width=0.328\textwidth]{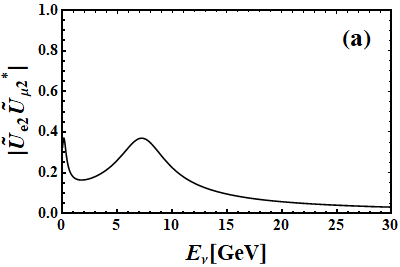}
  \includegraphics[width=0.328\textwidth]{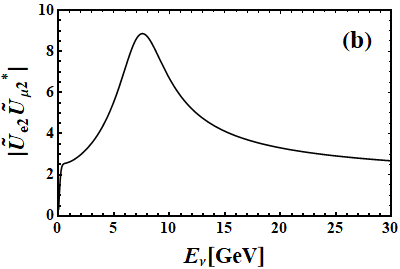}
  \includegraphics[width=0.328\textwidth]{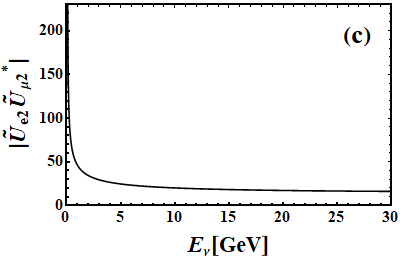}
  \caption{
     Plots of $\left| \tilde{U}_{e 2} \tilde{U}^*_{\mu 2}\right|$ with respect to the neutrino energy $E_\nu$ : (a) is obtained from eq.~\eqref{eq:mixUinmatdifab}, (b) is from ~\cite{Zhang:2006yq} and (c) is from~\cite{Li:2018ezt}.}
  \label{fig:3diffplts}
\end{figure}

Having discussed about behaviors of the matter-induced effective mass squares with energy and matter density, the last important ingredient of matter effect is the relation between the mixing matrix in vacuum and matter. One can find the relation by using both the unitarity condition of $\tilde{U}$ and taking up to the third power of a equation obtained by equating the right-hand sides of eqs.~\eqref{eq:effham1} and \eqref{eq:effham2}, which are
%%%%%
\be
\begin{cases}
\displaystyle \sum^4_i  \tilde{U}_{\alpha i} \tilde{U}^*_{\beta i} = \delta_{\alpha\beta}\,, \\[4mm]
\displaystyle \sum^4_i  \tilde{m}^2_i \tilde{U}_{\alpha i}\tilde{U}^*_{\beta i} = \sum^4_i  m^2_i U_{\alpha i} U^*_{\beta i} + \f_{\alpha \beta} \,, \\[4mm]
\displaystyle \sum^4_i  \tilde{m}^4_i \tilde{U}_{\alpha i}\tilde{U}^*_{\beta i} = \sum^4_i \left[ m^4_i  + m^2_i \left( \f_{\alpha \alpha} + \f_{\beta \beta}  \right)  \right] U_{\alpha i} U^*_{\beta i} + \f^2_{\alpha \beta}\,, \\[4mm]
\displaystyle \sum^4_i  \tilde{m}^6_i \tilde{U}_{\alpha i}\tilde{U}^*_{\beta i} = \sum^4_i \left[ m^6_i  + m^4_i \left( \f_{\alpha \alpha} + \f_{\beta \beta}  \right)  + m^2_i \left( \f^2_{\alpha \alpha} + \f^2_{\beta \beta} + \f_{\alpha \alpha} \f_{\beta \beta}   \right) \right] U_{\alpha i} U^*_{\beta i} \\[4mm] 
 \qquad \qquad \qquad \quad \displaystyle + \sum_{ij\gamma}m^2_i m^2_j U_{\alpha i} U^*_{\gamma i}U_{\gamma j}U^*_{\beta j}\f_{\gamma \gamma} + \f^3_{\alpha \beta}\,, 
\end{cases}
\label{eq:modtransforms}
\ee
%%%%%
where we defined $\f = 2 E \diag\left(V_\mathrm{CC}, 0, 0, -V_\mathrm{NC}\right)$. Clearly, these are the set of linear equations of the variables $\tilde{U}_{\alpha i} \tilde{U}^*_{\beta i}$, which can be solved in a rather straightforward way. As the right-hand side of the first equation becomes zero or one depending on whether or not the indices $\alpha$ and $\beta$ being equal, and $\f_{\alpha \beta}$ also takes different value accordingly, our solutions are written as the following
\begin{align}\label{eq:mixmatmodsqrt}
\left| \tilde{U}_{\alpha i} \right|^2 = & \left(\prod^4_{k\neq i}\D\tilde{m}^2_{ik} \right)^{-1} \left\{ \sum^4_{j=1} \left[ \prod^4_{r\neq i}  \left( \f_{\alpha \alpha} - \delta m^2_{rj}  \right) \right] \left| U_{\alpha j} \right|^2 \right. \notag \\
& \left. - \frac{1}{2} \sum_{m n\gamma} \left( \D m^2_{mn}\right)^2 U_{\alpha m} U^*_{\alpha n} U^*_{\gamma m} U_{\gamma n} \f_{\gamma \gamma} \right\} \, ,
\end{align}
and 
\begin{align}\label{eq:mixUinmatdifab}
\tilde{U}_{\alpha i} \tilde{U}^*_{\beta i} = & \left(\prod^4_{k\neq i}\D\tilde{m}^2_{ik} \right)^{-1} \left\{ \sum^4_{j=1} \left[ \prod^4_{r\neq i}  \left( \f_{\alpha \alpha} + \f_{\beta \beta} - \delta m^2_{rj}  \right) -\frac{3}{2}\left( \delta m^2_{ij}\right)^2 \left(  \f_{\alpha \alpha} + \f_{\beta \beta} \right)  \right. \right. \notag \\
& \left. - \delta m^2_{ij} \left[ \left(  \f_{\alpha \alpha} + \f_{\beta \beta} \right) \sum^4_{l=1} \D \tilde{m}^2_{l i} -2 \left(  \f_{\alpha \alpha} + \f_{\beta \beta} \right)^2 - \f_{\alpha \alpha}\f_{\beta \beta}  \right] \; \right] U_{\alpha j} U^*_{\beta j} \notag \\
& \left. - \frac{1}{2} \sum_{m n\gamma} \left( \D m^2_{mn}\right)^2 U_{\alpha m} U^*_{\beta n} U^*_{\gamma m} U_{\gamma n} \f_{\gamma \gamma} \right\} \, ,
\end{align}
for $\alpha \neq \beta$. Here we defined mass-squared differences $\D \tilde{m}^2_{ij} = \tilde{m}^2_i - \tilde{m}^2_j$,  $\D m^2_{ij} = m^2_i - m^2_j$ and $\delta m^2_{ij} = \tilde{m}^2_i - m^2_j$. Note that the form of eq.~\eqref{eq:mixUinmatdifab} does not match with the corresponding results given in~\cite{Zhang:2006yq,Li:2018ezt}, but we checked it in different ways and confirmed that this is a correct solution. First of all, our derivation of these solutions is given in appendix~\ref{app:sumrld4mixing}. Furthermore, as we will see in the below, these solutions return to the vacuum case when matter potentials become zero. Meanwhile, we did following numerical checks for this result: (i) we confirmed that this solution indeed satisfies all of the unitarity constraints, $\sum_i \tilde{U}_{\alpha i} \tilde{U}^*_{\beta i} =0 $ for all $\alpha \neq \beta$; (ii)we cross checked the eqs.~\eqref{eq:mixmatmodsqrt} and \eqref{eq:mixUinmatdifab} via two sides of the identity $\left| \tilde{U}_{\alpha i}\right|^2 \left| \tilde{U}_{\beta i}\right|^2 = \left| \tilde{U}_{\alpha i} \tilde{U}^*_{\beta i}\right|^2$ obtained from them; (iii) we also checked if $\left| \tilde{U}_{\alpha i} \tilde{U}^*_{\beta i}\right|\le 1$ holds for all $\alpha \neq \beta$ and $i=1, 2, 3, 4$ and found that this solution satisfies these conditions also. To exemplify this point and illustrate how different our result than those given in \cite{Zhang:2006yq} and \cite{Li:2018ezt}, here we show a plot of $\left| \tilde{U}_{e 2} \tilde{U}^*_{\mu 2}\right|$ in figure~\ref{fig:3diffplts}. When we draw these plots we use the parameters in table~\ref{tab:paramcollect} and choose $\delta_{14}=0$ and $\delta_{24}=0$ for simplicity. Last but not the least, the oscillation probability $P(\nu_\mu \to \nu_e)$ in matter, which we will discuss in the section~\ref{sec:nusrch@DUNE}, obtained from our solution confirms the result from simulation~\cite{Gandhi:2015xza}.

%
%%%%%%%%%%%%%%%%
%%%%%%%%%%%%%%%%
%

At first glance, results in eqs.~\eqref{eq:mixmatmodsqrt} and \eqref{eq:mixUinmatdifab} seem rather bewildering, but their meaning become clear if their forms are compactly thought as 
\be \label{eq:mixcompctfrm}
\tilde{U}_{\alpha i} \tilde{U}^*_{\beta i} = \sum^4_{j=1}C^{\alpha \beta}_{ij}  U_{\alpha j} U^*_{\beta j} + D^{\alpha \beta}_i ,
\ee
for both cases of  $\alpha$ and $\beta$ becoming equal or not.\footnote{Note that, here, we are referring to the forms of the solutions, but not to explicit definitions of these coefficients in each solution. } The coefficients $C^{\alpha \beta}_{ij}$ and $D^{\alpha \beta}_i$ are functions of neutrino energy, mass-squared differences and matter potentials, with the following properties: a) the $C^{\alpha \beta}_{ij}$ is real and symmetric for swapping the indices $\alpha$ and $\beta$, while the $D^{\alpha \beta}_i$ is hermitian, i.e. $D^{\alpha \beta}_i = \left(D^{\beta \alpha}_i\right)^*$; b) they satisfy initial conditions $C^{\alpha \beta}_{ij}(0) = \delta_{ij}$ and  $D^{\alpha \beta}_i(0)=0$, meaning that their values in the vacuum, which will be clear from the following discussions.

There are several comments in order for the results in eqs.~\eqref{eq:mixmatmodsqrt} and \eqref{eq:mixUinmatdifab}. First of all, these relations are independent of the parameterization of mixing matrix, since they connect the matrix elements but not the mixing angles. Of course, according to relations between the sines (or cosines) of the mixing angles and modulus of the mixing matrix entries, one can extract the relation between the mixing angles in matter and vacuum after a specific parameterization is chosen.  For instance, making use of the parameterization in eq.~\eqref{eq:U4mixmat}, expressions for the sines of the mixing angles in matter are
\begin{align}
 & \sin \tilde{\theta}_{14} = \left| \tilde{U}_{e4}\right|,&& \sin \tilde{\theta}_{13} =  \frac{\left| \tilde{U}_{e3}\right|}{\sqrt{1-\left| \tilde{U}_{e4}\right|^2}}\, , \notag \\
 & \sin \tilde{\theta}_{24} =  \frac{\left| \tilde{U}_{\mu 4}\right|}{\sqrt{1-\left| \tilde{U}_{e4}\right|^2}} , && \sin \tilde{\theta}_{12} = \frac{\left| \tilde{U}_{e2}\right|}{\sqrt{1- \left| \tilde{U}_{e3}\right|^2-\left| \tilde{U}_{e4}\right|^2}}\, , \notag \\
 & \sin \tilde{\theta}_{34} = \frac{\left| \tilde{U}_{\tau 4}\right| }{\sqrt{1- \left| \tilde{U}_{e4}\right|^2-\left| \tilde{U}_{\mu 4}\right|^2}}\, , 
\end{align}
and
\begin{equation*}
 \sin \tilde{\theta}_{23} = \left| \tilde{U}_{\mu 3}\right| \left[\left( \left| \tilde{U}_{e 1}\right|^2 + \left| \tilde{U}_{e 2}\right|^2 \right)\left( \left| \tilde{U}_{\tau 4}\right|^2 + \left| \tilde{U}_{s 4}\right|^2 \right)\right]^{-1/2}  \left| 1- \left| \tilde{U}_{e 4}\right|^2 + \frac{\left( \tilde{U}_{e 3} \tilde{U}^*_{\mu 3} \right) \left( \tilde{U}_{\mu 4} \tilde{U}^*_{e 4} \right) }{\left| \tilde{U}_{\mu 3}\right|^2} \right| \, .
\end{equation*}
Second, if some mixing matrix elements (or equivalently mixing angles) are zero in vacuum, they may be generated by the matter effect. This is easy to see from the compact form in eq.~\eqref{eq:mixcompctfrm} as it contains contributions from two terms of a rotation and a shift. Last but not the least, when matter density (or neutrino energy) goes to zero, these relations give vacuum results as they should. Specifically, when $\f = 0$, which corresponds to vacuum, parts in the curly brackets in eqs.~\eqref{eq:mixmatmodsqrt} and \eqref{eq:mixUinmatdifab} return to vacuum solution, because
\begin{align}
 & \left. \D\tilde{m}^2_{ij}\right|_{\f = 0} = \left. \delta m^2_{ij}\right|_{\f = 0}  = \D m^2_{ij}, \notag \\
 & \sum^4_{j=1} \left[\prod^4_{r\neq i}   \D m^2_{rj} \right] \left| U_{\alpha j} \right|^2 =   \left[ \prod^4_{k\neq i}   \D m^2_{ki} \right] \left| U_{\alpha i} \right|^2 ,\\
 & \sum^4_{j=1} \left[ \prod^4_{r\neq i}  \D m^2_{rj} \right] U_{\alpha j} U^*_{\beta j} \notag = \left[ \prod^4_{k\neq i}  \D m^2_{ki} \right] U_{\alpha i} U^*_{\beta i}\, .
\end{align}
This in turn derives the initial condition of $C^{\alpha \beta}_{ij}$ in the compact expression \eqref{eq:mixcompctfrm}.
In one of the following sections we will use these results to do phenomenological study in the long-baseline neutrino oscillation experiments.

\begin{figure}[ht]
  \centering
  \includegraphics[width=0.24\textwidth]{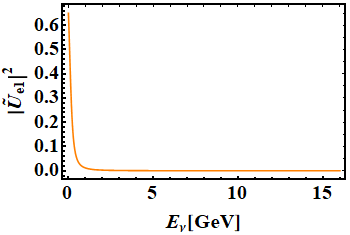}
  \includegraphics[width=0.24\textwidth]{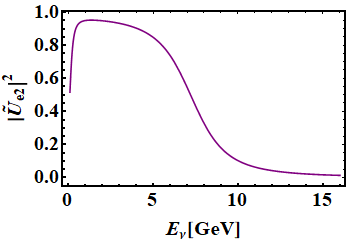}
  \includegraphics[width=0.245\textwidth]{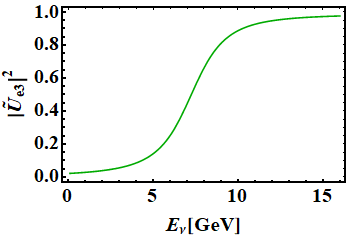}
  \includegraphics[width=0.245\textwidth]{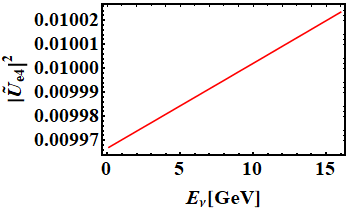}
  \includegraphics[width=0.24\textwidth]{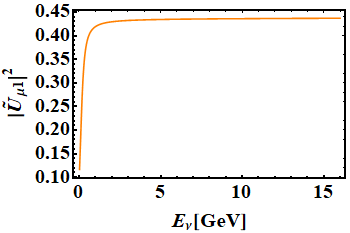}
  \includegraphics[width=0.24\textwidth]{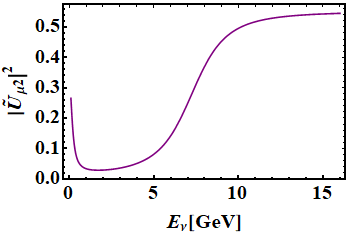}
  \includegraphics[width=0.245\textwidth]{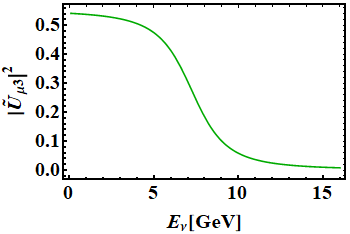}
  \includegraphics[width=0.245\textwidth]{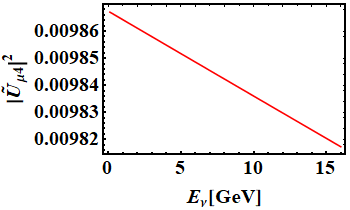}
  \includegraphics[width=0.24\textwidth]{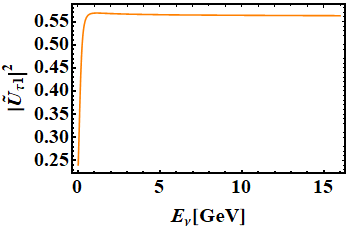}
  \includegraphics[width=0.24\textwidth]{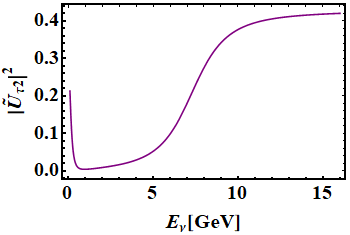}
  \includegraphics[width=0.245\textwidth]{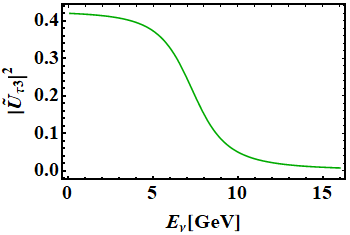}
  \includegraphics[width=0.245\textwidth]{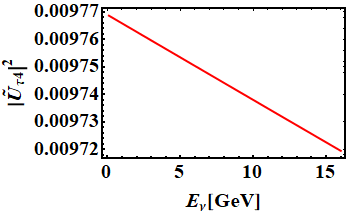}
  \includegraphics[width=0.24\textwidth]{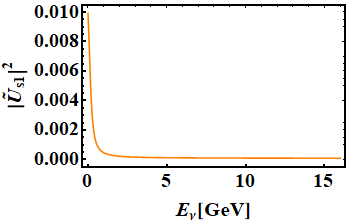}
  \includegraphics[width=0.24\textwidth]{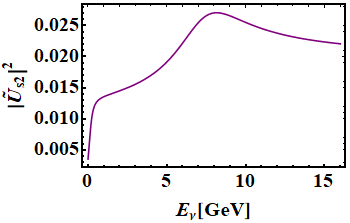}
  \includegraphics[width=0.245\textwidth]{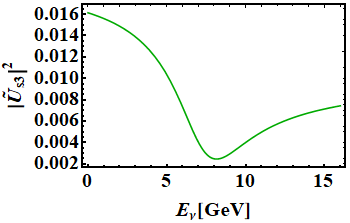}
  \includegraphics[width=0.245\textwidth]{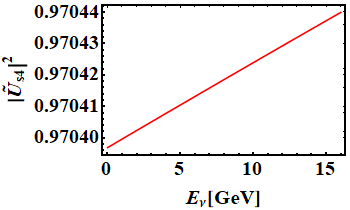}
  \caption{
    Shown are the plots of matter-effected mixing matrix elements' modulus squares with respect to the neutrino energy $E_\nu$.}
  \label{fig:Umodsqr}
\end{figure}

Figure~\ref{fig:Umodsqr} illustrates behaviors of matter-effected mixing matrix entries' modulus squares as functions of neutrino energy. When making these plots we assume that mixing angles $\theta_{24}$ and $\theta_{34}$  are of the same sizes as  $\theta_{14}$ and two Dirac CP-violating phases $\delta_{14}$ and $\delta_{24}$ are zero, while taking the best fit values of other active neutrino parameters in \cite{Esteban:2018azc} (see table~\ref{tab:paramcollect} for the complete list of parameter values). We noticed that changing the values of CP-violating phases $\delta_{14}$ and $\delta_{24}$ modify those curves slightly but qualitative behaviors do not change much. It can be seen from this figure that the modulus squares of the mixing matrix elements with active neutrino flavor indices and $i=1, 2, 3$ tend to reach some constant values at high energies.
%
%%%%%%%%%%%%%%%%
%%%%%%%%%%%%%%%%
%

Similar discussions can apply for the case of three active neutrinos and we obtain the following relations 
\begin{align}\label{eq:3numixmodsqrt}
\left| \tilde{U}_{\alpha i} \right|^2 =  \left(\prod^3_{k\neq i}\D\tilde{m}^2_{ik} \right)^{-1} \sum^3_{j=1} \left[ \prod^3_{r\neq i}  \left( \f^\prime_{\alpha \alpha} - \delta m^2_{rj}  \right) \right] \left| U_{\alpha j} \right|^2 \, ,
\end{align}
and
\begin{align}\label{eq:3mixmatdb}
\tilde{U}_{\alpha i} \tilde{U}^*_{\beta i} =  \left(\prod^3_{k\neq i}\D\tilde{m}^2_{ik} \right)^{-1}  \sum^3_{j=1} \left[ \prod^3_{r\neq i}  \left( \f^\prime_{\alpha \alpha} + \f^\prime_{\beta \beta} - \delta m^2_{rj}  \right) + \delta m^2_{ij} \left(  \f^\prime_{\alpha \alpha} + \f^\prime_{\beta \beta} \right)  \right] U_{\alpha j} U^*_{\beta j}  \, ,
\end{align}
where $\f^\prime \equiv 2 E \diag\left(V_\mathrm{CC}, 0, 0 \right)$, $i$ = 1, 2, 3 and $\alpha$, $\beta$ = $e$, $\mu$, $\tau$. These expressions have similar structures as the results given in eqs.~\eqref{eq:mixmatmodsqrt} and \eqref{eq:mixUinmatdifab}, so it is easy to repeat analogous comments we made in this section for three active neutrinos case.

%%%%%%%%%%%%%%%%%%%%%%%
\section{CP asymmetry with four neutrinos}
\label{sec:CPviolation}
%%%%%%%%%%%%%%%%%%%%%%%

In regard to the sources of CP asymmetry in four-neutrino oscillations, in vacuum and matter, one of the important consequences of introducing a light sterile neutrino is to generate several independent rephasing invariants, as a contrast to the only one such invariant with standard three neutrinos. As the CP transformation changes neutrino to its antineutrino and vice versa, and there is a relation between neutrino and antineutrino oscillation probabilities: $P_{\bar{\nu}_\alpha \to \bar{\nu}_\beta}(L,E)= P_{\nu_\beta \to \nu_\alpha}(L,E)$ or, in other words, $P_{\bar{\nu}_\alpha \to \bar{\nu}_\beta}= P_{\nu_\alpha \to \nu_\beta}\left(U \leftrightarrow U^*\right)$, an indicator of CP conservation in neutrino oscillation is to test if oscillation probabilities of neutrinos and antineutrinos are equal. Following this argument, a measure of CP asymmetry is
\be
\D P_{\alpha\beta} \equiv P_{\nu_\alpha \to \nu_\beta} - P_{\bar{\nu}_\alpha \to \bar{\nu}_\beta} = 4 \sum_{i < j}\im \left(U_{\alpha i}U_{\beta j}U_{\alpha j}^*U_{\beta i}^* \right)\sin\frac{\Delta m^2_{ji} L}{2E}\, .
\label{eq:cpaymP}
\ee
Form of this equation remains the same for both discussions in vacuum and matter, up to replacing the corresponding mixing matrix and mass-squared differences for the case under consideration. It is not hard to see that this quantity is antisymmetric with respect to the swapping of indices, which reduces the independent components to $n(n-1)/2$ when considering the $n$ neutrino flavors. What is more, applying the unitarity condition of mixing matrix,
\be
\sum_\alpha U_{\alpha i} U^*_{\alpha j} = \delta_{ij},
\ee
it is a very simple exercise to show that 
\be
\sum_\alpha \D P_{\alpha\beta} = 0 .
\label{eq:dpabsum}
\ee
This further imposes $n-1$ independent equations (not $n$ because one of them can always be obtained by using other $n-1$ constraints), so total number of independent $\D P_{\alpha\beta}$ is $n(n-1)/2 - (n-1) = (n-1)(n-2)/2$. Interestingly, this number coincides with the number of Dirac CP-violating phases with $n-1$ flavors. With this result in mind, one can easily obtain the number of independent components of $\D P_{\alpha\beta}$ for arbitrary number of flavors. For instance, there are 3 such quantities with 4 neutrinos and only one for the case of 3 neutrinos. Observing the expression \eqref{eq:cpaymP}, one may notice that imaginary part of the product of four mixing matrix entries can provide a rephasing invariant form, so CP asymmetry can also be quantified this way in terms of the Jarlskog invariants \cite{Jarlskog:1985ht,Wu:1985ea}
\be \label{eq:JIdefn}
J^{\alpha\beta}_{ij}\equiv \im \left(U_{\alpha i}U_{\beta j}U_{\alpha j}^*U_{\beta i}^* \right).
\ee
Definition above reveals that these quantities possess following properties
\be
J^{\alpha\beta}_{ij} = -J^{\beta\alpha}_{ij} = - J^{\alpha\beta}_{ji} = J^{\beta\alpha}_{ji}.
\label{eq:jkogprop1}
\ee
Furthermore, the unitarity of mixing matrix leads to 
\be
\sum_\alpha J^{\alpha\beta}_{ij} = \sum_i J^{\alpha\beta}_{ij} = 0 \, , 
\label{eq:jkogprop2}
\ee
through a similar argument to get eq.~\eqref{eq:dpabsum}. To sum up, just relying on the eq.~\eqref{eq:jkogprop1}, there are $\left[n(n-1)/2\right]^2$ independent rephasing invariants. Eq.~\eqref{eq:jkogprop2} further eliminates $n(n-1)^2/2 + (n-1)^2(n-2)/2$ of them, thus the total number of independent Jarlskog invariants in the case of $n$ flavors is $\left[(n-1)(n-2)/2\right]^2$. The same conclusion can be drawn from the fact that upper and lower indices are independent and each of them has $(n-1)(n-2)/2$ independent components. This is exactly a square of the number of independent components of $\D P_{\alpha\beta}$. A summary for the cases of some specific number of flavors is given in table~\ref{tab:indjkin}.
\begin{table}
\centering
\renewcommand{\arraystretch}{1.2}
\begin{tabular}{lcc|lcc}
  \toprule 
  $n \qquad$ & $\D P_{\alpha\beta}$ & $J^{\alpha\beta}_{ij}$ & $n\qquad$ & $\D P_{\alpha\beta}$ & $J^{\alpha\beta}_{ij}$ \\
  \midrule
  1 & 0 & 0 & 2 & 0 & 0 \\
  3 & 1 & 1 & 4 & 3 & 9 \\
  5 & 6 & 36 & 6 & 10 & 100 \\
  7 & 15 & 225 & 8 & 21 & 441 \\
  9 & 28 & 784 & 10 & 36 & 1296 \\ 
  \bottomrule
\end{tabular}
\caption{A summary of the total number of independent $\D P_{\alpha\beta}$ and Jarlskog invariants $J^{\alpha\beta}_{ij}$ correspond to number of neutrino flavors up to 10.}
\label{tab:indjkin}
\end{table}

Having discussed general case of $n$ family, the following discussions particularly focus on the Jarlskog invariants of four families. As we have argued, only nine of them are independent thus we can choose
\begin{align}\label{eq:jibasis}
& J^{e\mu}_{13}, \quad J^{e\mu}_{24}, \quad J^{e\mu}_{34}, \notag \\
& J^{e s}_{13}, \quad J^{e s}_{24}, \quad J^{e s}_{34} ,  \\
& J^{\mu s}_{13}, \quad J^{\mu s}_{24}, \quad J^{\mu s}_{34} , \notag 
\end{align}
as a basis.\footnote{Although any nine independent Jarlskog invariants can be chosen as basis, these are the ones with shorter expressions for a given parameterization in eq.~\eqref{eq:U4mixmat}.} All the other rephasing invariants are the linear combinations of these bases, explicit relations are given in table~\ref{tab:jinvrelsn}.
\begin{table}[ht]
\centering
\renewcommand{\arraystretch}{1.2}
\begin{tabular}{l|l|l}
  \toprule 
   $J^{e \mu}_{14} = -J^{e \mu}_{24} - J^{e \mu}_{34}$ & $J^{e \mu}_{12} = - J^{e \mu}_{13} - J^{e \mu}_{14}$ & $J^{e \mu}_{23} = -J^{e \mu}_{13} + J^{e \mu}_{34}$ \\
   $J^{e s}_{14} = -J^{e s}_{24} - J^{e s}_{34}$ & $J^{e s}_{12} = - J^{e s}_{13} - J^{e s}_{14}$ & $J^{e s}_{23} = -J^{e s}_{13} + J^{e s}_{34}$ \\
   $J^{\mu s}_{14} = -J^{\mu s}_{24} - J^{\mu s}_{34}$ & $J^{\mu s}_{12} = -J^{\mu s}_{13} - J^{\mu s}_{14}$ & $J^{\mu s}_{23} = - J^{\mu s}_{13} + J^{\mu s}_{34}$ \\
   $J^{e \tau}_{12} = -J^{e \mu}_{12} - J^{e s}_{12}$ & $J^{e \tau}_{13} = -J^{e \mu}_{13} - J^{e s}_{13}$ & $J^{e \tau}_{14} = -J^{e \mu}_{14} - J^{e s}_{14}$ \\
   $J^{e \tau}_{23} = -J^{e \mu}_{23} - J^{e s}_{23}$ & $J^{e \tau}_{24} = -J^{e \mu}_{24} - J^{e s}_{24}$ & $J^{e \tau}_{34} = -J^{e \mu}_{34} - J^{e s}_{34}$ \\
   $J^{\mu \tau}_{12} = J^{e \mu}_{12} - J^{\mu s}_{12}$ & $J^{\mu \tau}_{13} = J^{e \mu}_{13} - J^{\mu s}_{13}$ & $J^{\mu \tau}_{14} = J^{e \mu}_{14} - J^{\mu s}_{14}$ \\
   $J^{\mu \tau}_{23} = J^{e \mu}_{23} - J^{\mu s}_{23}$ & $J^{\mu \tau}_{24} = J^{e \mu}_{24} - J^{\mu s}_{24}$ & $J^{\mu \tau}_{34} = J^{e \mu}_{34} - J^{\mu s}_{34}$ \\
   $J^{\tau s}_{12} = J^{e \tau}_{12} + J^{\mu \tau}_{12}$ & $J^{\tau s}_{13} = J^{e \tau}_{13} + J^{\mu \tau}_{13}$ & $J^{\tau s}_{14} = J^{e \tau}_{14} + J^{\mu \tau}_{14}$ \\
%   \midrule
   $J^{\tau s}_{23} = J^{e \tau}_{23} + J^{\mu \tau}_{23}$ & $J^{\tau s}_{24} = J^{e \tau}_{24} + J^{\mu \tau}_{24}$ & $J^{\tau s}_{34} = J^{e \tau}_{34} + J^{\mu \tau}_{34}$\\
  \bottomrule
\end{tabular}
\caption{Expressions of other 27 Jarlskog invariants in terms of chosen bases in eq.~\eqref{eq:jibasis}.}
\label{tab:jinvrelsn}
\end{table}

To know CP asymmetry in four-neutrino oscillations, what has to be done is to show a dependence of these nine independent Jarlskog invariants in eq.~\eqref{eq:jibasis} on the mixing angles and Dirac CP-violating phases. So far, our calculations are independent of the parameterization chosen for mixing matrix. In order to see explicit expressions, one has to choose a basis and perform computations with definition in eq.~\eqref{eq:JIdefn}. Using the parameterization in eq.~\eqref{eq:U4mixmat}, expressions of these independent Jarlskog invariants in eq.~\eqref{eq:jibasis} are
\begin{align}\label{eq:Jinvs9}
J^{e\mu}_{24} = & s_{12} c_{13} s_{14} c^2_{14} s_{24} c_{24}\left[s_{12} s_{13} s_{23} \sin \left(\delta_{13}-\delta_{14}+\delta_{24}\right)+c_{12} c_{23} \sin \left(\delta_{14}-\delta_{24}\right)\right], \notag \\
J^{e s}_{24} = & s_{12} c_{13} s_{14} c^2_{14}c_{24} c_{34} \left\{c_{12} \left[\sin \delta_{14} s_{23} s_{34}-c_{23} s_{24} c_{34} \sin \left(\delta_{14}-\delta_{24}\right)\right] \right. \notag \\
& \left. -s_{12} s_{13} \left[s_{23} s_{24} c_{34} \sin \left( \delta_{13}-\delta_{14}+\delta_{24}\right)+c_{23} s_{34} \sin \left(\delta_{13}-\delta_{14}\right)\right]\right\}, \notag \\
J^{\mu s}_{24} = & \frac{1}{16} c^2_{14} s_{2(24)} c_{34} \left\{s_{34} \left[4 s_{14} s_{24} \left(s ^2_{12} s_{2 (13)} c_{23} \sin (\delta_{13}-\delta_{14})-\sin\delta_{14} s_{2(12)}c_{13} s_{23}\right) \right. \right. \notag  \\
& \left. +c_{24} \left(4 s_{2 (12)} s_{13} (\cos \delta_{13} \sin \delta_{24}c_{2(23)}-\sin \delta_{13} \cos \delta_{24})+\sin \delta_{24} s_{2(23)} \left(2 s^2_{12} c_{2(13)}+3 c_{2(12)}+1\right)\right)\right]   \notag \\
&\left.+4 s_{12} s_{14} c_{34} \left[s_{12} s_{2(13)} s_{23} \sin (\delta_{13}- \delta_{14}+ \delta_{24})+2 c_{12} c_{13} c_{23}\sin ( \delta_{14}- \delta_{24})\right]\right\}, \notag \\
J^{e \mu}_{34} = & -\frac{1}{8}s_{2(13)} s_{2(14)}s_{2(24)} s_{23} c_{14} \sin (\delta_{13}-\delta_{14}+\delta_{24}) , \notag \\
J^{e s}_{34} = & \frac{1}{4}s_{2(13)} s_{2(14)} c_{14} c_{24}c_{34} \left[s_{23} s_{24} c_{34} \sin (\delta_{13}-\delta_{14}+ \delta_{24})+c_{23} s_{34} \sin ( \delta_{13}- \delta_{14})\right],\notag \\
J^{\mu s}_{34} = &-\frac{1}{4} c^2_{14} \left[s_{2(13)}s_{14} s_{23} s_{2 (24)} c^2_{34} \sin (\delta_{13}-\delta_{14}+ \delta_{24})+\frac{1}{2}s_{2(24)} s_{2(34)} \left(s_{2(13)} s_{14} c_{23} s_{24} \sin ( \delta_{13}-\delta_{14})\right.\right.\notag \\
& \left.\left. +\sin \delta_{24} c^2_{13} s_{2 (23)} c_{24}\right)\right], \notag \\
J^{e\mu}_{13} = &-\frac{1}{4} c_{12} s_{2(13)} c^2_{14} \left[s_{14} s_{2 (24)} (c_{12} s_{23} \sin (\delta_{13}- \delta_{14}+ \delta_{24})- s_{12} s_{13} c_{23} \sin ( \delta_{14}-\delta_{24})) \right. \notag \\
& \left. +\sin  \delta_{13} s_{12} c_{13} s_{2(23)} c^2_{24}\right], \notag \\
J^{e s}_{13} = &\frac{1}{16} c_{12} s_{2(13)} c^2_{14} \left\{4 s_{14} \left[s_{2(24)} c^2_{34} (s_{12} s_{23} \sin ( \delta_{13}- \delta_{14}+ \delta_{24})-s_{12} s_{13} c_{23} \sin ( \delta_{14}- \delta_{24})) \right. \right. \notag \\
& \left. \left. + c_{24} s_{2(34)} (c_{12} c_{23} \sin ( \delta_{13}- \delta_{14})+\sin \delta_{14} s_{12} s_{13} s_{23})\right]+s_{12} c_{13} \left[4 s_{24} s_{2(34)} (\cos  \delta_{13}\sin \delta_{24} \right. \right. \notag \\
& \left. \left. -\sin \delta_{13} \cos  \delta_{24} c_{2(23)})+\sin  \delta_{13} s_{2(23)} \left((c_{2(24)}-3) c_{2(34)}+2  c^2_{24}\right)\right]\right\}
\end{align}
where $s_{ij} = \sin \theta_{ij}$, $c_{ij} = \cos \theta_{ij}$, $s_{n(ij)} = \sin (n\theta_{ij})$ and $c_{n(ij)} = \cos (n\theta_{ij})$. Because of its rather lengthy expression, the dependence of $J^{\mu s}_{13}$ on mixing angles and CP-violating phases is given in appendix~\ref{app:jcpexp}. It is easy to see that all of these rephasing invariants in eq.~\eqref{eq:Jinvs9} vanish if all CP-violating phases become zero. This is not only a check for the correctness of derivation but also necessary to preserve CP symmetry. Another set of linearly independent Jarlskog invariants obtained from a different parameterization of mixing matrix is given in~\cite{Guo:2001yt}, see also other relevant works in~\cite{Barger:1999hi,Kalliomaki:1999ii,Donini:1999he,Xing:2001bg}. 

Now that we have found the relation in eq.~\eqref{eq:mixcompctfrm}, writing the Jarlskog invariants in matter in terms of that in vacuum is not difficult
\begin{align}
\tilde{J}^{\alpha \beta}_{ij} = &\sum_{kl} C^{\alpha \beta}_{ik} C^{\alpha \beta}_{jl}J^{\alpha \beta}_{kl}- \sum_k \left[ C^{\alpha \beta}_{ik} \im \left( D^{\alpha \beta}_j U^*_{\alpha k} U_{\beta k} \right) - C^{\alpha \beta}_{jk} \im \left( D^{\alpha \beta}_i U^*_{\alpha k} U_{\beta k} \right)\right] \notag \\
&+ \im\left( D^{\alpha \beta}_i D^{\beta \alpha }_j\right),
\end{align}
where $C^{\alpha \beta}_{ij}$ and $D^{\alpha \beta}_i$ get their expressions from eq.~\eqref{eq:mixUinmatdifab}, since the Jarlskog invariants are identically zero in case $\alpha =\beta$ by definition. Moreover, finding a sum rule between the Jalskog invariants in matter and in vacuum is also rather easy. We start with the eq.~\eqref{eq:effham1} and eq.~\eqref{eq:effham2}, observing that off-diagonal entries are not effected by matter potential, that is
\be
\sum^4_{i=1}m^2_i U_{\alpha i} U^*_{\beta i} =\sum^4_{i=1}\tilde{m}^2_i \tilde{U}_{\alpha i} \tilde{U}^*_{\beta i}
\ee
when $\alpha \neq \beta$. Making use of the unitarity of mixing matrix, multiplying both sides by their complex conjugates and, then, taking imaginary parts lead to 
\be
\sum_{ij} \Delta m^2_{i1} \Delta m^2_{j2} J^{\alpha \beta}_{ij} = \sum_{ij} \Delta \tilde{m}^2_{i1} \Delta \tilde{m}^2_{j2} \tilde{J}^{\alpha \beta}_{ij}. 
\ee
Taking the same steps, one can also derive following several relations
\begin{align}
\sum_{ij} \Delta m^2_{i1} \Delta m^2_{j2} J^{\alpha \beta}_{ij} & = \sum_{ij} \delta m^2_{i1} \delta m^2_{j2} \tilde{J}^{\alpha \beta}_{ij} ,  \notag \\
\sum_{ij} \delta m^2_{1i}\,  \delta m^2_{2j} J^{\alpha \beta}_{ij} & = \sum_{ij} \Delta \tilde{m}^2_{i1}\, \Delta \tilde{m}^2_{j2} \tilde{J}^{\alpha \beta}_{ij}, \\
\sum_{ij} \delta m^2_{1i}\,  \delta m^2_{2j} J^{\alpha \beta}_{ij} & = \sum_{ij} \delta m^2_{i1}\, \delta m^2_{j2} \tilde{J}^{\alpha \beta}_{ij} \notag . 
\end{align}
In the expressions above, we used the first and second mass eigenstate indices, but one has freedom to use any other indices if needed and resulting expressions still remain correct. Namely, these sum rules still hold even when the index ``1" is replaced by $k$ and the index ``2" is replaced by $l$. Note that if the unsummed indices on the left-hand side become equal, then the sum vanishes; and the same happens to the expression on the right-hand side.

As far as the CP-violating observables are concerned,  there are three such independent quantities which can be chosen as $\D P_{\mu e}$, $\D P_{\mu \tau}$ and $\D P_{e \tau}$. Other such observables are the linear combinations of these three:
\begin{align}
\D P_{e s} & =\D P_{\mu e} - \D P_{e \tau}\, , \notag \\
\D P_{\mu s} & = -\D P_{\mu e} - \D P_{\mu \tau}\, ,  \\
\D P_{\tau s} & =\D P_{e \tau} + \D P_{\mu \tau}\, \notag .
\end{align}
Having found the relations of Jarlskog invariants in matter to that in vacuum, we can easily obtain the relation between the matter-affected CP asymmetry $\D \tilde{P}_{\alpha \beta}$ and its counterpart in vacuum from the expression below
\be
\D \tilde{P}_{\alpha \beta} = 4 \sum_{i<j} \tilde{J}^{\alpha \beta}_{ij} \sin \frac{\D \tilde{m}^2_{ji} L}{2E} .
\ee
Experimental signature of these quantities will be discussed in the next section, focusing, in particular, on proposed long-baseline neutrino oscillation experiments.

%%%%%%%%%%%%%%%%%%%%%%%
\section{Implications for long-baseline experiments }
\label{sec:nusrch@DUNE}
%%%%%%%%%%%%%%%%%%%%%%%

 Heretofore we have discussed effects of sterile neutrino, purely on the theoretical ground, to three active neutrino scheme, including oscillations in vacuum and matter as well as the paradigm change in CP asymmetry with this additional neutrino flavor. To show its phenomenological consequence, in this section we will try to  embody possible hints in long-baseline neutrino oscillation experiments such as DUNE, NO$\nu$A and T2HK.

Deep Underground Neutrino Experiment (DUNE)~\cite{Acciarri:2016crz,Acciarri:2015uup,Strait:2016mof,Acciarri:2016ooe} is a world-leading long-baseline neutrino experiment planned to operate near future with 40 kiloton liquid argon detector at the Sanford underground research facility, located 1300 km downstream of the source at the Fermi national laboratory. This is a multipurpose experiment to unveil some neutrino-related mysteries in particle physics, such as determination of the mass ordering, measurement of the CP-violating phase in lepton sector, pin down the octant of $\theta_{23}$, search for a new physics beyond three-neutrino paradigm, precision measurement of neutrino parameters and many others. As stated previously, one of the primary goal of DUNE experiment is to search for a new physics focusing on the precision measurement of the parameters in muon neutrino and muon antineutrino oscillation channels.

The NO$\nu$A (NuMI Off-axis $\nu_e$ Appearance) experiment~\cite{Ayres:2007tu} is another long-baseline experiment that aims to measure oscillations of $\nu_\mu \to \nu_e$ in one of its 14 kiloton detector (made up of liquid scintillators contained in PVC) located 810 km away from neutrino source at Fermilab. The main goal of NO$\nu$A includes the precision measurement of atmospheric mixing angle, mass-squared differences and also put constraints on CP-violating phase.

T2HK (Tokai-to-Hyper-Kamiokande)~\cite{Abe:2015zbg} is an extension of T2K experiment, which uses water Cherenkov detector of 1 megaton  volume that to be placed about 295 km away from the source of neutrino beam at J-PARC (Japan  Proton  Accelerator  Research  Complex). The main purpose of T2HK is to study CP asymmetry in the lepton  sector using accelerator neutrino and anti-neutrino beams. 

%%%%%%%%%%%%%%%%
\begin{figure}%[ht]
  \centering
  \includegraphics[width=0.325\textwidth]{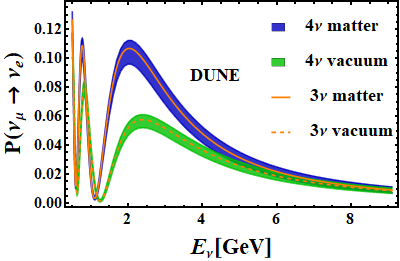}
  \includegraphics[width=0.325\textwidth]{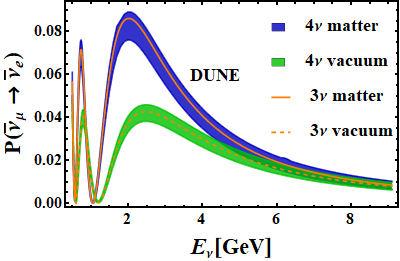} 
  \includegraphics[width=0.325\textwidth]{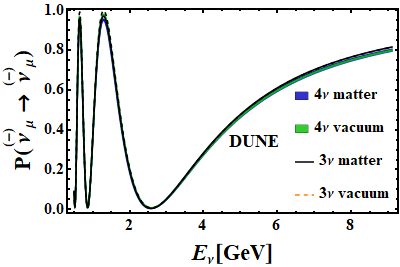} 
  \includegraphics[width=0.325\textwidth]{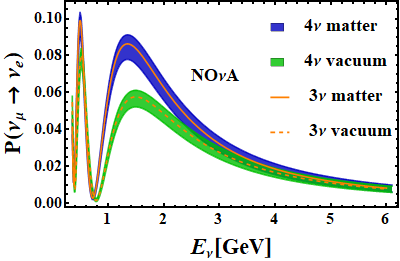}
  \includegraphics[width=0.325\textwidth]{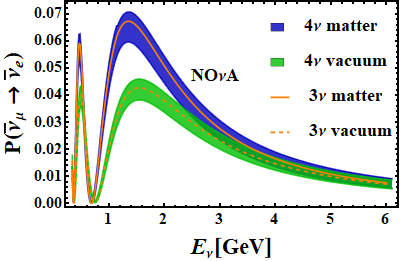} 
  \includegraphics[width=0.325\textwidth]{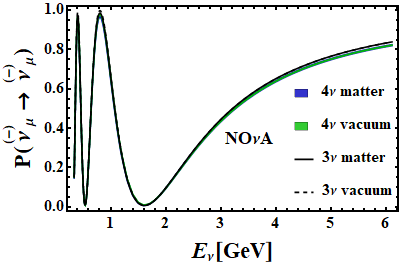}
  \includegraphics[width=0.325\textwidth]{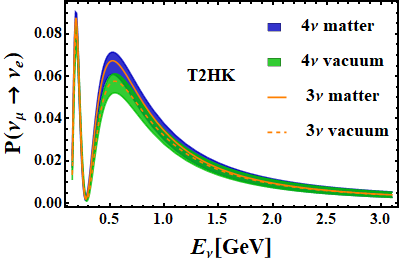}
  \includegraphics[width=0.325\textwidth]{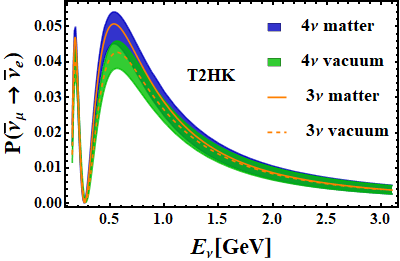} 
  \includegraphics[width=0.325\textwidth]{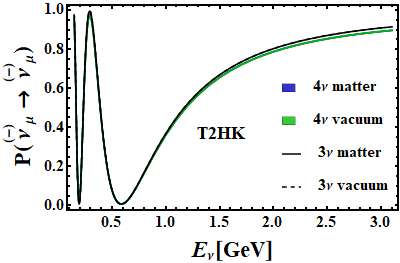}
  \caption{
    Prospects of muon neutrino and antineutrino oscillation probabilities in DUNE, NO$\nu$A and T2HK experiments, listed from the top to bottom}
  \label{fig:DUNEpros}
\end{figure}
%%%%%%%%%%%%%%%%
%

To illustrate our results, throughout the paper we assume that the $1\, \mathrm{eV}$ mass sterile neutrino has a percent level mixing with all active neutrinos. This is a suitable assumption because it is in the best-fit allowed region of the recent global fit results. As Refs.~\cite{Boser:2019rta,Diaz:2019fwt} have reported, from the global fit analysis, that the best-fit values of sterile neutrino parameters are $\D m^2_{41} \simeq 1.3 \, \mathrm{eV}^2$, $\left| U_{e 4} \right|^2\simeq0.012$, and $\left| U_{\mu 4} \right|^2\simeq0.018$. The values/ranges of neutrino oscillation parameters used in our analysis are listed in the table~\ref{tab:paramcollect}.
According to the values of mass-squared differences in this table, one can see that there is a hierarchical pattern $\Delta m^2_{4i}\gg \Delta m^2_{31}, \Delta m^2_{21}$ (for $i=1, 2, 3$) and thus expect that there are very rapid oscillations driven by $\Delta m^2_{4i}$ in both neutrino oscillation probability and CP asymmetry curves of 3+1 case. We investigate this expectation and confirm there are indeed such fast oscillations that these long-baseline experiments hardly reach sensitivities to observe them. So in our subsequent discussions we average out these modes by equating $\sin^2\frac{\Delta m^2_{4i} L}{4E}$ with $\frac{1}{2}$ and $\sin\frac{\Delta m^2_{4i} L}{2E}$ with $0$ (for $i=1, 2, 3$) in eq.~\eqref{eq:nuosprob},
\begin{equation}\label{eq:nuosproblttties}
\begin{aligned} 
P\left(\nu_\alpha \to \nu_\beta \right) = \ &\delta_{\alpha\beta}\left[ 1-2\re \left( U^*_{\alpha 4} U_{\beta 4}\right) \right] + 2 \left|  U_{\alpha 4}\right|^2  \left|  U_{\beta 4}\right|^2\\
& -4\sum^3_{i < j}\re \left(U_{\alpha i}U_{\alpha j}^*U_{\beta i}^*U_{\beta j}\right)\sin^2\frac{\Delta m^2_{ij} L}{4E} \\
& - 2\sum^3_{i < j}\im \left(U_{\alpha i}U_{\alpha j}^*U_{\beta i}^*U_{\beta j}\right)\sin\frac{\Delta m^2_{ij} L}{2E}\, .
\end{aligned}
\end{equation}
In particular,
\be
\begin{aligned} 
P\left({\nu_\mu \to \nu_e}\right) = \ & 2 \left|  U_{\mu 4}\right|^2  \left|  U_{e 4}\right|^2 -4\sum^3_{i < j}\re \left(U_{\mu i}U_{\mu j}^*U_{e i}^*U_{e j}\right)\sin^2\frac{\Delta m^2_{ij} L}{4E} \\
& - 2\sum^3_{i < j}\im \left(U_{\mu i}U_{\mu j}^*U_{e i}^*U_{e j}\right)\sin\frac{\Delta m^2_{ij} L}{2E}\,,
\end{aligned}
\ee
and 
\be
P\left({\nu_\mu \to \nu_\mu}\right) = \ 1- 2 \left( \left|  U_{\mu 4}\right|^2 -  \left|  U_{\mu 4}\right|^4 \right) - 4\sum^3_{i < j} \left|  U_{\mu i}\right|^2 \left|  U_{\mu j}\right|^2 \sin^2\frac{\Delta m^2_{ij} L}{4E}\, .
\ee
When discussing the antineutrino oscillation probabilities, one just needs to change signs of the terms including imaginary part.  The same averaging is also implemented for CP asymmetry in the 3+1 neutrino scheme,
\be
\D P_{\alpha \beta} = -4 \sum^3_{i<j} \im \left(U_{\alpha i}U_{\alpha j}^*U_{\beta i}^*U_{\beta j}\right) \sin \frac{\D m^2_{ij} L}{2E} .
\ee
Again, these expressions can apply to both cases in vacuum and matter by replacing the corresponding oscillation parameters. Based on our exact analytic expressions and averaging out of high frequency modes, we display both muon neutrino and antineutrino oscillation probabilities in figure~\ref{fig:DUNEpros}. 
%%%%%%%%%%%%%%%%%%%%%%%%%%%%%
\begin{table}%[ht]
\centering
\renewcommand{\arraystretch}{1.8}
\begin{tabular}{|c|c|}
  \hline
  parameters & values/ranges \\ \hline
   $\sin^2 \theta_{12}$ & $0.310$ \\ \hline
   $\sin^2 \theta_{23}$ & $0.558$ \\ \hline
   $\sin^2 \theta_{13}$ & $0.022$ \\ \hline
   $\sin^2 \theta_{i4}$ & $0.010$ \\ \hline
   $\delta_{13}/\pi$ & $1.19$ \\ \hline
   $\delta_{14}/\pi$ & $[0,2]$ \\ \hline
   $\delta_{24}/\pi$ & $[0,2]$ \\ \hline
   $\frac{\Delta m^2_{21}}{10^{-5}~\mathrm{eV}^2}$ & $7.39$ \\ \hline
   $\frac{\Delta m^2_{31}}{10^{-3}~\mathrm{eV}^2}$ & $2.53$ \\ \hline
   $\frac{\Delta m^2_{41}}{1~\mathrm{eV}^2}$ & $1.0$ \\ \hline
\end{tabular}
\caption{List of neutrino oscillation parameter values (or ranges) used in our phenomenological study.}
\label{tab:paramcollect}
\end{table}
%%%%%%%%%%%%%%%%%%%%%%%%%%%%%%%%
As a reminder, although the averaging out of the fourth mass contribution makes oscillation curve of $3+1$ case a single line, blue and green bands in these plots are due to the allowing the two CP-violating phases $\delta_{14}$ and $\delta_{24}$ to change in the interval $[0,2\pi]$.  According to figure~\ref{fig:DUNEpros}, it is clear to see that in the appearance channels of $\overset{\scriptscriptstyle(-)}{\nu}_e$ there are rather significant separations in the probabilities for both cases of three- and four-neutrino oscillations, thanks to the matter effect. Important energy ranges to find these separations are 1 -- 4 GeV in DUNE, 0.8 -- 2.2 GeV in NO$\nu$A and 0.4 -- 0.7 GeV in T2HK. All of these energy ranges are under the coverage of typical energy of neutrinos from accelerator. Although the oscillation probabilities between matter and vacuum are well separated, three active neutrino oscillation curves pass through the bands of four-neutrino oscillation probabilities. Displacement between the oscillation curves with and without the sterile neutrino may (or may not) be apparent when the values of $\delta_{14}$ and $\delta_{24}$ become precise. Another aspect we learn from this figure is that the separation between the matter and vacuum oscillation probabilities gets bigger as the baseline increases. Therefore, knowing the values of $\delta_{14}$ and $\delta_{24}$ and increasing the baseline of an experiment are necessary to make a claim about the detection of active-sterile neutrino oscillations. As long as these two Dirac CP-violating phases measured precise enough and a sizable separation is confirmed to be there, it might more likely for DUNE to distinguish 3+1 neutrino oscillation signal from the three active neutrino background. As a further step, an experiment with much longer baseline and with high sensitivity could probably distinguish four-neutrino oscillation from three active neutrinos case with the help of matter effect; and the eV-scale sterile neutrino signal could be detected by looking at the appearance mode of $\overset{\scriptscriptstyle(-)}{\nu}_e$. The disappearance channel of $\overset{\scriptscriptstyle(-)}{\nu}_\mu$, however, cannot distinguish the sterile neutrino signal from the background of active neutrino oscillations even after measuring the values of Dirac CP-violating phase. That neither has noticeable dependence on these two phases nor has pronounced change over the baseline of the experiments. For this reason, a would-be golden channel to find active-sterile neutrino oscillation is $\overset{\scriptscriptstyle(-)}{\nu}_\mu \to \overset{\scriptscriptstyle(-)}{\nu}_e$.\footnote{We investigate, for completeness, $\overset{\scriptscriptstyle(-)}{\nu}_\mu \to \overset{\scriptscriptstyle(-)}{\nu}_\tau$ oscillation channel and find that it is also not very effective way to look for the sterile neutrino contribution. In addition, since this channel is not of direct use for any of the long-baseline experiments discussed in this paper, we do not show these plots here.}

Comparing the results from $\overset{\scriptscriptstyle(-)}{\nu}_\mu \to \overset{\scriptscriptstyle(-)}{\nu}_e$ oscillation channels, the DUNE has a greater potential to observe larger oscillation probability and bigger separation enjoyed by the larger matter effect, due to its longer baseline. Neutrino energies correspond to the first pick value of oscillation probabilities, at which point the separation between the oscillation in matter and vacuum is most significant, moves from 2 GeV in the DUNE to 1.4 GeV in NO$\nu$A and to 0.5 GeV in T2HK. And, at the same time, not only do these pick values decrease but also the gap between the oscillation curves in matter and vacuum shrinks with the decrease of baseline. 

\begin{figure}[!h]
  \centering
  \includegraphics[width=0.6\textwidth]{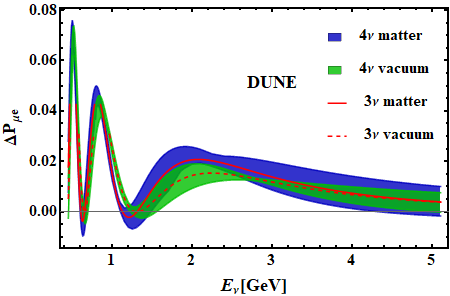}
  \includegraphics[width=0.6\textwidth]{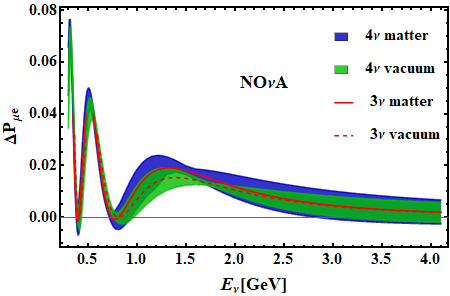}
  \includegraphics[width=0.6\textwidth]{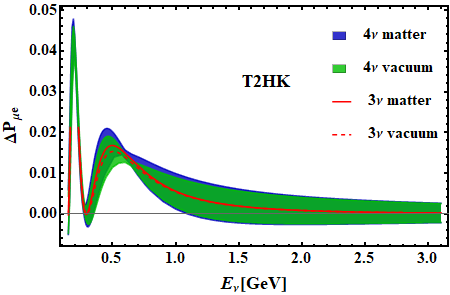}
  \caption{
    Shown are the plots of CP asymmetries in the long-baseline experiments DUNE, NO$\nu$A and T2HK with respect to the neutrino energy, and comparison between  the results in 3+0 and 3+1 neutrino schemes.}
  \label{fig:DUNECPV}
\end{figure}

As for the CP asymmetry in neutrino oscillations, figure~\ref{fig:DUNECPV} illustrates both neutrino energy and two new CP-violating phase dependence of  $\D P_{\mu e}$ and $\D \tilde{P}_{\mu e}$ in the long-baseline experiments that are under consideration. Blue and green bands are the allowed regions from varying both $\delta_{14}$ and $\delta_{24}$ in their full ranges. As this figure shows, 3+0 CP asymmetry lines pass through the allowed regions of 3+1 case. And, although a sizable amount of CP asymmetry appeared when neutrino energy less than 4.0 GeV for DUNE, 2.6 GeV for NO$\nu$A and 1.0 GeV for T2HK, no enhancement or clear separation can be seen from the matter effect as well as the presence of the sterile neutrino. So, it might be very hard, or not even possible, for these experiments to reach such a high precision to extract the sterile neutrino effects from the measurements of the CP asymmetry. High-energy tails of all CP asymmetry curves are damping to zero, thus, there is no hope to find CP asymmetry from this energy range either. Looking at the effects of two Dirac CP-violating phases $\delta_{14}$ and $\delta_{24}$, given the lack of their measured values by the current experimental data, it is hard to single out sterile neutrino contribution unless 3+1 CP asymmetry curves turn out to be deviated from 3+0 lines by staying close to upper or lower edges of the allowed regions.  When it comes to the comparison of the CP asymmetry curves for these three experiments, there is comparatively larger CP asymmetry in DUNE than T2HK, as it is illustrated that the longer baseline the experiment has, the larger value of $\D P_{\mu e}$ can arise. But separations between the $\D \tilde{P}_{\mu e}$ curves with and without the sterile neutrino is not clear. Although all of these long-baseline experiments could be able to measure sizable CP asymmetry, there might be a big challenge for them to aim at finding the light sterile neutrino signal in the CP asymmetry measurements.

%%%%%%%%%%%%%%%%
%

%%%%%%%%%%%%%%%%%%%%%%%%%%%%%%%%%
\section{Summary and conclusion}
\label{sec:conclusions}
%%%%%%%%%%%%%%%%%%%%%%%%%%%%%%%%%

The work in this paper is a theoretical study about the possibility of searching for a light sterile neutrino and observing the CP asymmetry in the long-baseline neutrino oscillations. To this end, we have discussed matter effects to neutrino masses, mixing and oscillation probabilities, by providing their relations in matter and vacuum. We have also carried out a thorough analysis on CP-violating quantities by detailed derivation of independent Jarlskog invariants, CP asymmetries, and their connection in matter to vacuum. Based on our results, we have performed  phenomenological study of four-neutrino oscillations, in comparison with the standard three-neutrino framework, in DUNE, NO$\nu$A and T2HK experiments and presented the CP asymmetries from the relevant oscillation channels.  

Our analysis indicate that more promising ways to search for the imprints of the light sterile neutrino in these experiments are focusing on the electron neutrino and electron antineutrino appearance channels. As expected, the matter effects play a non-trivial role in searching for the sterile neutrino, by amplifying quite significantly the oscillation probabilities. Figure~\ref{fig:DUNEpros} illustrates that the separation between the oscillation probability in matter and vacuum increase as the increase of the neutrino propagation distance in these experiments. We also notice that sizes of the oscillation probabilities get larger with the longer baseline. But the separations between the probability curves of four-neutrino signal from the three active neutrino background is also subject to the values of CP-violating phases. In the oscillation curves for DUNE, in contrast to other two experiments, there are slightly bigger probability and lager separation when comparing the cases with and without the matter effect. Such a  prominent separation in the electron (anti)neutrino appearance channel arises within the energy range from 1 GeV to 4 GeV, while no similar phenomenon occurs in the muon (anti)neutrino disappearance channels. This kind of deviations, and the energy ranges they are appearing, in the NO$\nu$A and T2HK get smaller with the decrease of the baseline. 

From the CP asymmetry plots in figure~\ref{fig:DUNECPV} one can see, unlike the situation in oscillation probability, that there are no such clear distinctions between the cases whether or not taking into account of the matter effect. Moreover, three-neutrino CP asymmetry curves in vacuum and matter are respectively within the corresponding allowed regions of four-neutrino cases. This means that it might be challenging for these experiments to reach such high precision to extract sterile neutrino contribution, unless the values of CP-violating phases make the separation between the results from four- and three-neutrino cases large enough.  Comparing the results for these three experiments, there is no big difference among them. Nonetheless, size of the CP asymmetry and value of the neutrino energy up to which noticeable CP asymmetry can arise increase with the longer baseline. So, in this sense, DUNE could observe a bit bigger value of CP asymmetry and scan larger range of neutrino energy than that in the other two experiments.

Finally, as stated, the results in this paper based on the theoretical study of long-baseline neutrino oscillations and CP asymmetries. These results could anticipate what would be the outcome, without considering the experimental capabilities, but they are not enough to claim full discovery potentials of these experiments. In order to make a complete statement one must consider also other experimental factors like neutrino flux, cross section, detector size, energy resolution, uncertainty, and efficiency etc, which are left to be discussed in a separate work. 

%%%%%%%%%%%%%%%%%%%%%%%
\section*{Acknowledgements}
%%%%%%%%%%%%%%%%%%%%%%%
We are grateful to Xiaoyong Chu for useful comments and suggestions on the manuscript. This work was supported by the National Natural Science Foundation of China under the Grant No. 11875306.

%%%%%%%%%%%%%%%%%%%%%%%
\appendix
%%%%%%%%%%%%%%%%%%%%%%%

%%%%%%%%%%%%%%%%%%%%%%%%%%%%%%%%%
\section{Matter-induced effective mass squares}
\label{app:sol4polyeq}
%%%%%%%%%%%%%%%%%%%%%%%%%%%%%%%%%

This section is to calculate effective mass squares from the equation
\be
(\tilde{m}_i^2)^4 - c_3 (\tilde{m}_i^2)^3 + c_2 (\tilde{m}_i^2)^2 - c_1 \tilde{m}_i^2  + c_0 =0 ,
\label{eq:quarticsol}
\ee
where $c_3, c_2, c_1, c_0$ are the trace, the sum of determinants of $2\times 2$ main diagonal blocks, the sum of determinants of $3\times 3$ main diagonal blocks and the determinant of $2EH_\mathrm{eff}$. Results from our calculation are the following:
\begin{align}
 c_3 = & \Tr \left( 2EH_\mathrm{eff} \right) = \sum^4_i m^2_i + 2E \left(V_\mathrm{CC} - V_\mathrm{NC}\right) , \notag \\
 c_2 = & \sum^4_{i< j} m^2_i m^2_j + 2EV_\mathrm{CC} \sum^4_i m^2_i \left( 1 - \left|U_{ei}\right|^2 \right) - 2EV_\mathrm{NC} \sum^4_i m^2_i \left( 1 - \left|U_{si}\right|^2 \right) - 4E^2 V_\mathrm{CC}V_\mathrm{NC} , \notag \\
 c_1 = & \sum^4_{n,i,j,k} \varepsilon^2_{nijk} m^2_i m^2_j\left[ \frac{1}{6} m^2_k +  \left( V_\mathrm{CC} \left|      U_{ek}\right|^2 - V_\mathrm{NC} \left|U_{sk}\right|^2 \right) E \right] \notag \\
     & - 4E^2 V_\mathrm{CC}V_\mathrm{NC}\sum^4_i m^2_i \left( 1 - \left|U_{ei}\right|^2 - \left|U_{si}\right|^2\right) , \notag \\
 c_0 = & \ m^2_1  m^2_2  m^2_3  m^2_4 + \frac{E}{3}\sum^4_{n,i,j,k} \varepsilon^2_{nijk} m^2_i m^2_j m^2_k \left( V_\mathrm{CC} \left|U_{en}\right|^2 - V_\mathrm{NC} \left|U_{sn}\right|^2 \right) \notag \\
       & - E^2 V_\mathrm{CC} V_\mathrm{NC} \sum^4_{i,j,k, l} \varepsilon^2_{ijkl} m^2_i m^2_j  \left| U_{ek} U_{sl} - U_{el} U_{sk}  \right|^2 .
\end{align}
As one can see, all of those coefficients are non-negative real numbers.
To make life simple, let us write eq.~\eqref{eq:quarticsol} with $x\equiv \tilde{m}_i^2$ and then perform change of variable $x = y + \frac{c_3}{4}$. This provides
\be
y^4 +\left( -\frac{3c^2_3}{8}+c_2\right) y^2 + \left( -\frac{c^3_3}{8}+\frac{1}{2}c_3 c_2 - c_1\right) y + \left( -\frac{3c^4_3}{256} + \frac{c^2_3 c_2}{16} - \frac{1}{4} c_3 c_1 + c_0\right) = 0 \, .
\ee
To avoid carrying cumbersome coefficients, one can introduce the following variables
\begin{align}
 p & = \frac{8c_2 - 3c^2_3}{8} , \notag \\
 q & = -\frac{c^3_3 - 4c_3 c_2 + 8 c_1}{8} , \notag \\
 r & = \frac{-3c^4_3 +16 c^2_3 c_2 -64  c_3 c_1 + 256 c_0}{256} \, .
\end{align}
Using these new variables and adding $p^2/4 + 2z (y^2 + p/2)+ z^2 $ on the both sides, above equation can be transformed into the following form: 
\be
\left( y^2 + \frac{p}{2} + z \right)^2 = 2z y^2 - q y + z^2 + p z -r + \frac{p^2}{4}\, .
\label{eq:qsepou}
\ee
Since this equality holds for any values of $z$, one can choose its value to make the right-hand site full square. This can be done when the discriminant of quadratic equation of $y$ is zero. Because in this case the quadratic equation has two identical roots or, equivalently, the right-hand side of eq.~\eqref{eq:qsepou} becomes a complete square if and only if the discriminant vanishes, i.e. 
\be
q^2 - 8z\left( z^2 + p z -r + \frac{p^2}{4} \right) = 0 \, .
\ee
This, in turn, requires that $z$ satisfies cubic equation
\be
z^3 + pz^2 + \frac{1}{4}\left( p^2 - 4r \right) z - \frac{1}{8} q^2 =0 \, .
\label{eq:3bic}
\ee
As a result, solution of the quartic equation boils down to the solution of one quadratic equation,
\be
\left( y^2 + \frac{p}{2} + z \right)^2 - \left( \sqrt{2z} y - \frac{q}{2\sqrt{2z} } \right)^2 = 0 \, ,
\label{eq:qua2equns}
\ee
and one cubic equation in \eqref{eq:3bic}.
Solutions of the quadratic equation of $y$ in \eqref{eq:qua2equns} are easily obtained from
\be
y^2 + \sqrt{2z} y  + \frac{p}{2} + z - \frac{q}{2\sqrt{2z}} = 0 \, .
\ee
or
\be
y^2 - \sqrt{2z} y  + \frac{p}{2} + z + \frac{q}{2\sqrt{2z}} = 0
\ee
These two equations yield following four possible solutions of quartic equation
\begin{align}
 y_1 & = - \frac{1}{2} \left[ \sqrt{2z} + \sqrt{-2z - 2p +  \sqrt{\frac{2}{z} } \, q \;}  \; \right] , \notag \\
 y_2 & = - \frac{1}{2} \left[ \sqrt{2z} - \sqrt{-2z - 2p +  \sqrt{\frac{2}{z} } \, q \;}  \; \right] , \notag \\
 y_3 & =  \frac{1}{2} \left[ \sqrt{2z} - \sqrt{-2z - 2p -  \sqrt{\frac{2}{z} } \, q \;}  \; \right] , \notag \\
 y_4 & =  \frac{1}{2} \left[ \sqrt{2z} + \sqrt{-2z - 2p -  \sqrt{\frac{2}{z} } \, q \;}  \; \right] \, .
 \label{eq:4ticsol}
\end{align}
But this is not the end of story, to have complete solution, cubic equation of $z$ in \eqref{eq:3bic} must be solved. This will be done in the following steps. Since the solutions of interest are real ones, although there are many other ways to solve this equation, the trigonometric solution of cubic equation provides expected result. By doing a change of variable $z = t - \frac{1}{3}p$, it is easy to get read of the quadratic term in~\eqref{eq:3bic}, i.e.
\be
t^3 + vt + w= 0, 
\ee
where
\begin{align*}
 v & = -\frac{1}{12} p^2 - r \, , \notag \\
 w & = \frac{1}{216}\left( -2 p^3 - 27 q^2 + 72 p r \right) \, .
\end{align*}
Let us implement another replacement $t = u \cos \theta$, then divide both sides by $u^3/4$, the cubic equation above reads
\be
4 \cos^3 \theta + \frac{4 v}{u^2} \cos \theta + \frac{4w}{u^3} = 0 \; , 
\label{eq:tri3bicidtt}
\ee
which has very similar form as the well-known trigonometric identity
\be
4 \cos^3 \theta -3 \cos \theta - \cos 3\theta = 0 \; . 
\ee
Comparing these two equations, one can easily identify $u = 2\sqrt{-v/3}$ and replacing it back to eq.~\eqref{eq:tri3bicidtt} yields
\be
4 \cos^3 \theta -3 \cos \theta + \frac{3w}{2\left| v \right|} \sqrt{-\frac{3}{ v }} = 0 \; . 
\ee
This implies
\be
\theta = \frac{1}{3} \left[ \arccos \left( \frac{-3w}{2\left| v \right|} \sqrt{-\frac{3}{ v }}  \; \right) + 2k\pi \right] ,
\ee
for $k = 0, 1, 2$. Thus, solutions of cubic equation \eqref{eq:3bic} have the following form
\be
z_{k+1} = -\frac{1}{3}p + 2\sqrt{\frac{-v}{3}} \cos \left[ \frac{1}{3} \arccos \left( \frac{-3w}{2\left| v \right|} \sqrt{-\frac{3}{ v }} \; \right) + \frac{2k\pi}{3} \right] .
\ee
Finally, inserting this result to the eq.~\eqref{eq:4ticsol}, the solution for quartic equation comes to the end. Combining all we have got so far, solutions of eq.~\eqref{eq:quarticsol} are 
\begin{align}
\tilde{m}_1^2 & = \frac{c_3}{4} - \frac{1}{2} \left[ \sqrt{2z} + \sqrt{-2z - 2p +  \sqrt{\frac{2}{z} } \, q \;}  \; \right] , \notag \\
\tilde{m}_2^2 & = \frac{c_3}{4} - \frac{1}{2} \left[ \sqrt{2z} - \sqrt{-2z - 2p +  \sqrt{\frac{2}{z} } \, q \;}  \; \right] , \notag \\
\tilde{m}_3^2 & = \frac{c_3}{4} +  \frac{1}{2} \left[ \sqrt{2z} - \sqrt{-2z - 2p -  \sqrt{\frac{2}{z} } \, q \;}  \; \right] , \notag \\
\tilde{m}_4^2 & = \frac{c_3}{4} + \frac{1}{2} \left[ \sqrt{2z} + \sqrt{-2z - 2p -  \sqrt{\frac{2}{z} } \, q \;}  \; \right] \, .
% \label{eq:4ticsolfin}
\end{align}

%%%%%%%%%%%%%%%%%%%%%%%%%%%%%%%%%
\section{Derivation of the relations between mixing matrix in matter and vacuum}
\label{app:sumrld4mixing}
%%%%%%%%%%%%%%%%%%%%%%%%%%%%%%%%%
This part is dedicated to find relations between combinations of mixing matrix elements in matter and vacuum. The main purpose is to solve following linear equations system
\be
\begin{cases}
\displaystyle \sum^4_i  \tilde{U}_{\alpha i} \tilde{U}^*_{\beta i} = \delta_{\alpha\beta}\,, \\[4mm]
\displaystyle \sum^4_i  \tilde{m}^2_i \tilde{U}_{\alpha i}\tilde{U}^*_{\beta i} = \sum^4_i  m^2_i U_{\alpha i} U^*_{\beta i} + \f_{\alpha \beta} \,, \\[4mm]
\displaystyle \sum^4_i  \tilde{m}^4_i \tilde{U}_{\alpha i}\tilde{U}^*_{\beta i} = \sum^4_i \left[ m^4_i  + m^2_i \left( \f_{\alpha \alpha} + \f_{\beta \beta}  \right)  \right] U_{\alpha i} U^*_{\beta i} + \f^2_{\alpha \beta}\,, \\[4mm]
\displaystyle \sum^4_i  \tilde{m}^6_i \tilde{U}_{\alpha i}\tilde{U}^*_{\beta i} = \sum^4_i \left[ m^6_i  + m^4_i \left( \f_{\alpha \alpha} + \f_{\beta \beta}  \right)  + m^2_i \left( \f^2_{\alpha \alpha} + \f^2_{\beta \beta} + \f_{\alpha \alpha} \f_{\beta \beta}   \right) \right] U_{\alpha i} U^*_{\beta i} \\[4mm] 
 \qquad \qquad \qquad \quad \displaystyle + \sum_{ij\gamma}m^2_i m^2_j U_{\alpha i} U^*_{\gamma i}U_{\gamma j}U^*_{\beta j}\f_{\gamma \gamma} + \f^3_{\alpha \beta}\, . 
\end{cases}
\label{eq:modtransforms1}
\ee
When $\alpha = \beta$ this system of equations can be written in the following linear matrix equation form
\be
\mathbf{A} \mathbf{V} = \mathbf{B}\
\label{eq:mateqsys}
\ee
with
\be
\mathbf{A} = \begin{pmatrix} 
            1 & 1 & 1 & 1 \\
            \tilde{m}^2_1 & \tilde{m}^2_2 & \tilde{m}^2_3 & \tilde{m}^2_4  \\ \;
            \tilde{m}^4_1 & \tilde{m}^4_2 & \tilde{m}^4_3 & \tilde{m}^4_4  \\
            \tilde{m}^6_1 & \tilde{m}^6_2 & \tilde{m}^6_3 & \tilde{m}^6_4 
           \end{pmatrix} \, ,
\ee

\be
\mathbf{V} = \begin{pmatrix} \;
           \left| \tilde{U}_{\alpha 1}\right|^2 \; \; \\ \; 
           \left| \tilde{U}_{\alpha 2}\right|^2 \;  \; \\ \; 
           \left| \tilde{U}_{\alpha 3}\right|^2 \; \;  \\ \; 
           \left| \tilde{U}_{\alpha 4}\right|^2 \; \;
           \end{pmatrix} \, ,
\ee
and
\be
\mathbf{B} = \begin{pmatrix} 
           1 \\
      \displaystyle     \sum^4_i  m^2_i \left| U_{\alpha i} \right|^2 + \f_{\alpha \alpha} \  \\
      \displaystyle    \sum^4_i \left( m^4_i  + 2m^2_i \f_{\alpha \alpha}  \right) \left| U_{\alpha i} \right|^2 + \f^2_{\alpha \alpha} \\
      \displaystyle    \sum^4_i \left( m^6_i  + 2 m^4_i \f_{\alpha \alpha} + 3 m^2_i \f^2_{\alpha \alpha} \right) \left| U_{\alpha i} \right|^2
       + \sum_{ij\gamma}m^2_i m^2_j U_{\alpha i} U^*_{\alpha j} U^*_{\gamma i}U_{\gamma j} \f_{\gamma \gamma} + \f^3_{\alpha \alpha} \end{pmatrix} \, .
\ee
Interestingly, the matrix $\mathbf{A}$ is nothing but the Vandermonde matrix with elements $\mathbf{A}_{ij} = \tilde{m}^{2(i-1)}_j$ , which has well-known determinant 
\be
\det (\mathbf{A}) = \prod^4_{1\le i < j} \left( \tilde{m}^2_j -\tilde{m}^2_i \right) = \prod^4_{ i < j} \D \tilde{m}^2_{ji}  \, ,
\ee
and inverse 
\begin{align}
\displaystyle
\mathbf{A}^{-1} = & \begin{pmatrix} 
 \displaystyle                \prod_{ j \neq 1} \D \tilde{m}^2_{j1} &  &  &  \\
                  &   \displaystyle  \prod_{ j \neq 2} \D \tilde{m}^2_{j2} &  & \\                   &  &  &  \displaystyle \prod_{ j \neq 3} \D \tilde{m}^2_{j3} &  \\                   &  &  &   &  \displaystyle \prod_{ j \neq 4} \D \tilde{m}^2_{j4} \\[4mm]
                \end{pmatrix}^{-1} \notag \\
               &  \begin{pmatrix} 
                \tilde{m}^2_2 \tilde{m}^2_3 \tilde{m}^2_4 & - \tilde{m}^2_3 \tilde{m}^2_4 - \tilde{m}^2_2  \left( \tilde{m}^2_3 + \tilde{m}^2_4 \right)  & \tilde{m}^2_2 + \tilde{m}^2_3 + \tilde{m}^2_4 & -1  \\ 
               \tilde{m}^2_1 \tilde{m}^2_3 \tilde{m}^2_4 & - \tilde{m}^2_3 \tilde{m}^2_4 - \tilde{m}^2_1  \left( \tilde{m}^2_3 + \tilde{m}^2_4 \right) & \tilde{m}^2_1 + \tilde{m}^2_3 + \tilde{m}^2_4 & -1  \\ 
                \tilde{m}^2_1 \tilde{m}^2_2 \tilde{m}^2_4 & - \tilde{m}^2_2 \tilde{m}^2_4 - \tilde{m}^2_1  \left( \tilde{m}^2_2 + \tilde{m}^2_4 \right) & \tilde{m}^2_1 + \tilde{m}^2_2 + \tilde{m}^2_4 & -1   \\ 
                \tilde{m}^2_1 \tilde{m}^2_2 \tilde{m}^2_3 & - \tilde{m}^2_2 \tilde{m}^2_3 - \tilde{m}^2_1  \left( \tilde{m}^2_2 + \tilde{m}^2_3 \right) & \tilde{m}^2_1 + \tilde{m}^2_2 + \tilde{m}^2_3 & -1 
                \end{pmatrix}.
\end{align}
Finding solutions of linear equations system \eqref{eq:mateqsys} boils down to the computation of the following matrix product
\be
\mathbf{V} = \mathbf{A}^{-1} \mathbf{B}.
\ee
After a bit of computational work, one reaches to the solution
\begin{align}
\left| \tilde{U}_{\alpha i} \right|^2 = & \left(\prod^4_{k\neq i}\D\tilde{m}^2_{ik} \right)^{-1} \left\{ \sum^4_{j=1} \left[ \prod^4_{r\neq i}  \left( \f_{\alpha \alpha} - \delta m^2_{rj}  \right) \right] \left| U_{\alpha j} \right|^2 \right. \notag \\
& \left. - \frac{1}{2} \sum_{m n\gamma} \left( \D m^2_{mn}\right)^2 U_{\alpha m} U^*_{\alpha n} U^*_{\gamma m} U_{\gamma n} \f_{\gamma \gamma} \right\} \, .
\end{align}

On the other hand, when $\alpha \neq \beta$, discussion is the same as above except for replacing $\mathbf{V}$ by $\mathbf{V'}$ and $\mathbf{B}$ by $\mathbf{B'}$, which are
\be
\mathbf{V'} = \begin{pmatrix} \;
           \tilde{U}_{\alpha 1} \tilde{U}^*_{\beta 1} \; \; \\ \; 
           \tilde{U}_{\alpha 2} \tilde{U}^*_{\beta 2} \;  \; \\ \; 
           \tilde{U}_{\alpha 3} \tilde{U}^*_{\beta 3} \; \;  \\ \; 
           \tilde{U}_{\alpha 4} \tilde{U}^*_{\beta 4} \; \;
           \end{pmatrix} \, ,
\ee
\be
\mathbf{B'} = \begin{pmatrix} 
           0 \\
      \displaystyle     \sum^4_i  m^2_i U_{\alpha i} U^*_{\beta i}   \  \\
      \displaystyle    \sum^4_i \left[ m^4_i  + m^2_i \left( \f_{\alpha \alpha} + \f_{\beta \beta}  \right)  \right] U_{\alpha i} U^*_{\beta i}  \\
      \displaystyle    \sum^4_i \left[ m^6_i  + m^4_i \left( \f_{\alpha \alpha} + \f_{\beta \beta}  \right)  + m^2_i \left( \f^2_{\alpha \alpha} + \f^2_{\beta \beta} + \f_{\alpha \alpha} \f_{\beta \beta}   \right) \right] U_{\alpha i} U^*_{\beta i} \\
      \displaystyle + \sum_{ij\gamma}m^2_i m^2_j U_{\alpha i}U^*_{\beta j} U^*_{\gamma i}U_{\gamma j}\f_{\gamma \gamma} 
            \end{pmatrix} \, .
\ee
Now that the matrix $\mathbf{A}$ remains unchanged, the solutions are given by
\be
\mathbf{V'} = \mathbf{A}^{-1} \mathbf{B'}.
\ee
It is straightforward to compute matrix multiplications on the right-hand side of above equation. The explicit form of the solutions are
\begin{align}
\tilde{U}_{\alpha i} \tilde{U}^*_{\beta i} = & \left(\prod^4_{k\neq i}\D\tilde{m}^2_{ik} \right)^{-1} \left\{ \sum^4_{j=1} \left[ \prod^4_{r\neq i}  \left( \f_{\alpha \alpha} + \f_{\beta \beta} - \delta m^2_{rj}  \right) -\frac{3}{2}\left( \delta m^2_{ij}\right)^2 \left(  \f_{\alpha \alpha} + \f_{\beta \beta} \right)  \right. \right. \notag \\
& \left. - \delta m^2_{ij} \left[ \left(  \f_{\alpha \alpha} + \f_{\beta \beta} \right) \sum^4_{l=1} \D \tilde{m}^2_{l i} -2 \left(  \f_{\alpha \alpha} + \f_{\beta \beta} \right)^2 - \f_{\alpha \alpha}\f_{\beta \beta}  \right] \; \right] U_{\alpha j} U^*_{\beta j} \notag \\
& \left. - \frac{1}{2} \sum_{m n\gamma} \left( \D m^2_{mn}\right)^2 U_{\alpha m} U^*_{\beta n} U^*_{\gamma m} U_{\gamma n} \f_{\gamma \gamma} \right\} \, ,
\end{align}
where $\D \tilde{m}^2_{ij} \equiv \tilde{m}^2_i - \tilde{m}^2_j$,  $\D m^2_{ij} \equiv m^2_i - m^2_j$ and $\delta m^2_{ij} \equiv \tilde{m}^2_i - m^2_j$\,.

At this point, we obtained not only the matter-effected mixing matrix element's modulus squares but also the paired product of one element with complex conjugate of the other, which are essential to investigate matter effects in many neutrino physics processes including four neutrino oscillation as well as  CP asymmetry.

%%%%%%%%%%%%%%%%%%%%%%%%%%%%%%%%%
\section{Expression of the $J^{\mu s}_{13}$}
\label{app:jcpexp}
%%%%%%%%%%%%%%%%%%%%%%%%%%%%%%%%%

This section shows the expression of $J^{\mu s}_{13}$ as a function of mixing angles and Dirac CP-violating phases. This is the last piece of nine independent Jarlskog invariants analyzed in main body of the paper.
\begin{align*} 
J^{\mu s}_{13}= &  -\frac{1}{8}s_{2(12)} s_{2(13)}s_{2(23)}c_{13} c^2_{34} s^2_{14} s^4_{24} \sin\delta_{13} \\
& +\frac{1}{32} c^2_{34} s_{14} \left\{\left[32 c^3_{13} \sin \delta_{14} s_{13} s_{23} \left(\sin (\delta_{13}- \delta_{14}) \sin ( \delta_{14}- \delta_{24})s^2_{14}+\cos ( \delta_{13}+ \delta_{24}) s^2_{23}\right) s_{24} c^3_{24}  \right. \right. \\
& +4 s^2_{24} \left(c_{2(14)} c_{24} \sin \delta_{13} \sin ( \delta_{13}- \delta_{14}) \sin \delta_{14} s^2_{2(13)} s_{14}-\left(\cos ( \delta_{14}+ \delta_{24}) \sin \delta_{13} s_{4(13)} s^2_{14} \right. \right.\\
& \left. \left.  +2 \left(\cos ( \delta_{14}- \delta_{24}) \sin\delta_{13} c^2_{14}+\cos \delta_{13} (\cos \delta_{14} \sin \delta_{24}-\cos \delta_{24} c_{2(14)} \sin \delta_{14})\right) s_{2(13)}\right) s_{23} s_{24}\right) c_{24} \\
& +4 s_{2(13)} \left(2 c_{2(23)} \sin ( \delta_{13}- \delta_{14}+ \delta_{24}) s_{23} s^3_{24}-4 c^2_{24} \sin ( \delta_{13}- \delta_{14}) s^2_{13} s_{23} \left(\sin  \delta_{14} \sin (2  \delta_{13} - \delta_{14} \right. \right.\\
& \left. \left. + \delta_{24}) s^2_{14}+\cos \delta_{24} s^2_{23}\right) s_{24}+c_{24} \left(c_{2(24)}-c_{2(23)} c^2_{24}\right) \sin \delta_{13} \sin ( \delta_{13}- \delta_{14}) \sin \delta_{14} s_{2(13)} s_{14}\right) c_{24}  \\
& +((3 \cos ( \delta_{13}-2 \delta_{14})-\cos ( \delta_{13}+2 \delta_{24})+2 \cos ( \delta_{13}- \delta_{14}+ \delta_{24}) \cos ( \delta_{14}+ \delta_{24})c_{4(13)}) \sin  \delta_{13}- \\
& \left. 4 \cos \delta_{13} c_{2(13)} \sin ( \delta_{13}-2 \delta_{14})) s_{14} s^2_{23} s^2_{2(24)}\right] c^2_{12} + 8 c_{13} c_{23} s_{2(24)} s_{2(12)} \left(\cos  \delta_{24} c^2_{13} c^2_{24} \sin \delta_{14} s^2_{23} \right. \\
& \left. + \cos ( \delta_{13}- \delta_{14}+ \delta_{24}) \sin ( \delta_{13}- \delta_{14}) \sin ( \delta_{14}- \delta_{24}) \sin  \delta_{24} s ^2_{13} s^2_{14} s^2_{24}\right) + 2 s_{13}s_{14} \left(2 c^2_{23} \sin 2 ( \delta_{13}  \right. \\
& \left.  - \delta_{14}) s^2_{12}s_{13}-\cos (2 \delta_{24}) c^2_{13} \sin \delta_{13} s_{2(12)} s_{2(23)}\right) s^2_{2(24)} 
 +4 c_{23} s_{2(24)} \left[c_{13} s_{2(12)} \left(\left(\left(\cos ( \delta_{14}- \delta_{24}) \right. \right. \right. \right.   \\
& \left. \sin 2( \delta_{13}- \delta_{14})-2 \cos ^2 \delta_{24} \sin ( \delta_{14}- \delta_{24})\right) s^2_{13} s^2_{14}+2 \left(\sin ( \delta_{14}- \delta_{24})-\sin (2 \delta_{13}- \delta_{14}+ \delta_{24}) \right. \\
& \left. \left. \left.
 s^2_{13}\right)  s^2_{23}\right) s^2_{24}-2 c^2_{24} \sin ( \delta_{13}- \delta_{14}) s^2_{13} \left(\sin \delta_{14} \sin ( \delta_{13}- \delta_{14}+ \delta_{24}) s^2_{14}+2 \cos \delta_{13} \cos \delta_{24} s^2_{23}\right)\right) \\
&\left. \left.
-s^2_{12} s_{2(13)} s_{2(23)} \left(\cos \delta_{24} \sin ( \delta_{13}- \delta_{14}) c^2_{24}+\sin (\delta_{13}- \delta_{14}+ \delta_{24}) s^2_{24}\right)\right]\right\} +\\
& c_{13} s_{34} \left\{  \frac{1}{2} c_{12} s_{2(13)} c_{34} s^2_{14} s_{23} \left[  \cos \delta_{24} (\cos \delta_{13} - \cos (\delta_{13}-2\delta_{14}+2\delta_{24})) c_{12} c_{23} \sin \delta_{13} s_{13} + \sin ( \delta_{13} \right. \right.\\
&\left. \left.
 + \delta_{24})s_{12} s_{23} \right]s^3_{24}  + c_{23} \left[ \frac{1}{2}s_{2(12)} s_{2(13)}\sin\delta_{13} s_{23}(s^2_{14}s^2_{24} -c^2_{24}) + \frac{1}{8} s_{14} \left( s_{2(12)}  (2\sin(\delta_{13} - \delta_{14} + \right. \right. \right. \\
& \left. \left. \left. \left.  \delta_{24} )s^2_{13} s^2_{23} + (c_{2(23)} - c_{2(13)} ) \sin(\delta_{14} - \delta_{24})\right)
-2c_{2(12)} \sin (\delta_{13} -\delta_{14} + \delta_{24} ) s_{13} s_{2(23)} \right) s_{2(24)}\right]s_{34}\right\} \\
&  +\frac{1}{8} c_{13}s_{14} \left\{ 4 c_{34} \left[c_{24}s_{12} s_{23} \left(2 c_{12} \left(\sin ( \delta_{13}- \delta_{14}+ \delta_{24}) \left(\sin ( \delta_{13}- \delta_{14}) \sin ( \delta_{14}- \delta_{24}) s^2_{13} s^2_{14} \right. \right. \right. \right. \right. \\
&  \left. \left. +\cos ( \delta_{13}+ \delta_{24}) s^2_{23}\right)-\cos ( \delta_{13}- \delta_{14}+ \delta_{24}) c_{2(13)} \sin ( \delta_{13}+ \delta_{24}) s^2_{23}\right)+\sin ( \delta_{13}- \delta_{14}+2  \delta_{24}) \\
&
\left.  \left.  
s_{12} s_{13} s_{2(23)}\right)-c_{12} c_{23} s_{14} \left(\cos  \delta_{24} c_{12} c_{13} \sin (2  \delta_{13}) s_{23} s^2_{13}+c_{23} \sin ( \delta_{13}- \delta_{24}) s_{12} s_{2(13)}\right) s_{24}\right] s_{34}  \\
& 
+s_{13} \left(\cos \delta_{24} \cos ( \delta_{13}-2  \delta_{14}+2  \delta_{24}) c^2_{12} \sin  \delta_{13} s_{2(13)} s_{14} s_{2(23)} s_{24}-c_{24} s_{23} \left(4 \cos  \delta_{24} \sin ( \delta_{13}- \right. \right. \\
& \left. \left. \left. 
\delta_{14}+ \delta_{24}) s_{2(23)} c^2_{12}+(\cos \delta_{13}+\cos ( \delta_{13}-2 \delta_{14}+2  \delta_{24})) \sin ( \delta_{13}- \delta_{14}) s_{2(12)} s_{13} s^2_{14}\right)\right) s_{2(34)}\right\} s^2_{24} \\
&+\frac{1}{16}  c_{24} s_{2(34)} \left\{ c_{23} s_{14} \left[\cos  \delta_{14} c^2_{24} \sin \delta_{13} s_{4(13)} s^2_{23}+8 c_{24} \left(\cos ( \delta_{14}- \delta_{24}) \sin \delta_{14} c^4_{13}-\cos ( \delta_{13}+  \right.  \right.  \right. \\
& \left. 
\delta_{24}) \sin \delta_{13}s^2_{13} c^2_{13}+\cos ( \delta_{13}-\delta_{14}+ \delta_{24}) \sin ( \delta_{13}- \delta_{14}) s^4_{13}\right) s_{14} s_{24}s_{23}+4 s_{2(13)} \left(\cos \delta_{14} \sin \delta_{13} s^2_{13}  \right. \\
&\left. \left. 
- \cos \delta_{13} \sin \delta_{14}\right) s^2_{14} s^2_{24}\right] c^2_{12}+ 2 c_{24} s_{2(12)} \left[c_{13}c_{24} \left(2 c^2_{13} \sin \delta_{14} c^2_{23}+\cos \delta_{13} \sin ( \delta_{13}- \delta_{14}) s^2_{13}\right) \right.\\
& 
s_{14} s_{23}  +2 s_{13} \left(\left(\sin ( \delta_{13}- \delta_{14}) (\cos \delta_{13} \cos ( \delta_{13}- \delta_{14}+ \delta_{24}) c_{2(23)}+\sin \delta_{13} \sin ( \delta_{13}- \delta_{14}+ \delta_{24})) s^2_{13} \right. \right. \\
& \left. 
-\cos ( \delta_{13}- \delta_{14}+ \delta_{24}) c^2_{13} c^2_{23} \sin \delta_{14}\right) s^2_{14}+ c^2_{13} \left(\cos ( \delta_{14}- \delta_{24}) \sin ( \delta_{13}- \delta_{14}) s^2_{14}-\sin ( \delta_{13}+  \right. \\
&\left. \left. \left. 
\delta_{24})\right) s^2_{23}\right) s_{24}\right]  +2 c_{13} c_{23} s_{14} \left[\sin ( \delta_{14}-2  \delta_{24}) s_{2(12)} s_{2(23)}-2 \sin ( \delta_{13}- \delta_{14}) s_{13} \left(2 c^2_{23}s^2_{12}+ \right. \right. \\
& \left. \left. 
\cos  \delta_{13} s_{2(12)} s_{13} s_{2(23)}\right)\right] s^2_{24}+ \frac{1}{2} s_{2(13)} c^2_{24} s_{14} [( \sin \delta_{13}\cos (\delta_{14}) (1-3 c_{2(12)})+4 \cos \delta_{13} c_{2(12)} \\
&
\sin \delta_{14}) s_{23}s_{2(23)}-2 \cos \delta_{13} \sin ( \delta_{13}- \delta_{14}) s_{2(12)} s_{13} s_{3(23)}]-2s_{2(23)}s_{2(24)} s^2_{12} \left[\sin \delta_{24} c^2_{13}+\cos ( \delta_{13} \right. \\
& \left. \left.
- \delta_{14}+ \delta_{24}) \sin ( \delta_{13}- \delta_{14}) s^2_{13} s^2_{14}\right]  \right\} \\
&
 +\frac{1}{4}   c_{24} c_{34}  s_{14} \left\{ 4 \cos \delta_{14} c_{12} c_{13} \left[-\frac{1}{2}c_{12} s_{2(24)} s_{14} s_{23}  (c_{34} \sin \delta_{14} s_{23} s_{24} +c_{23} \sin (\delta_{14}- \delta_{24}) s_{34}) c^3_{13}  \right. \right. \\
& 
 +\left( c_{34} s_{23} s_{24} \left( c_{12} s_{13} \left(\left(\sin ( \delta_{13}-2 \delta_{14}+ \delta_{24}) s^2_{14} -\sin ( \delta_{13}+ \delta_{24}) s^2_{23}\right) c^2_{24}+\sin ( \delta_{13}+ \delta_{24}) s^2_{14} s^2_{24}\right) \right. \right.\\
 & %\left. \left. 
 -\frac{1}{2} s_{2(23)} c^2_{24} \sin \delta_{24} s_{12} ) +\frac{1}{2} c_{12}  c_{23} \sin  \delta_{13} s_{13} \left(c_{2(23)} c^2_{24} - c_{2(14)} s^2_{24} - c_{2(24)}\right) s_{34} ) c^2_{13}  \\
 &  
 +\frac{1}{8} c_{24} s_{2(13)} s_{14} \left( c_{34} \left(2 \sin ( \delta_{13}- \delta_{14}) \left(2 \cos \delta_{13} c_{12} s_{13} s^2_{23} +s_{12} s_{2(23)}\right) c^2_{24}+\left( c_{12} s_{13} \left(4 \sin (2  \delta_{13} \right. \right. \right. \right.   \\
 & \left. 
 - \delta_{14}+2  \delta_{24}) s^2_{23} -\sin (2  \delta_{13}- \delta_{14})-2 c_{2(14)} \sin  \delta_{14} +\sin  \delta_{14} \right)+2 \sin ( \delta_{13}-  \delta_{14}+2  \delta_{24}) s_{12} \\
 &\left. \left.
 s_{2(23)}\right) s^2_{24} +2 \cos ( \delta_{13}- \delta_{14}) c_{12} \left( c_{2(23)} c^2_{24} - c_{2(24)} \right) \sin \delta_{13} s_{13} \right)+2 \left(2 \sin (\delta_{13}- \delta_{14}+ \delta_{24}) s_{12} \right. \\
 & \left. \left. 
 c^2_{23} + c_{12} \sin (2  \delta_{13}- \delta_{14}+ \delta_{24}) s_{13} s_{2(23)} \right) s_{24} s_{34} \right) + c^2_{24} c_{34} s^2_{13} s^2_{14}  ( c_{23} \sin \delta_{24} s_{12} + c_{12} \sin ( \delta_{13}  \\
 & \left. 
 + \delta_{24}) s_{13} s_{23} ) s_{24} \right]+\cos ( \delta_{13}- \delta_{14}) s_{13} \left[2 \cos ( \delta_{14}- \delta_{24}) c^2_{12} s_{2(24)} c_{24} c_{34} \sin \delta_{14} s^2_{14}  s_{23} c^3_{13} +\frac{1}{4} c_{24} \right. \\
 & 
 c_{34} s_{14} \left(4 (\cos \delta_{24} \sin ( \delta_{14}- \delta_{24})+\cos ( \delta_{14}- \delta_{24}) c_{2(24)} \sin \delta_{24} ) s_{2(12)} s_{2(23)} -16 c^2_{12} s_{13}  \left(\cos \delta_{13} \right. \right. \\
 & \left. \left. 
 \sin \delta_{14} s^2_{14} +\cos ( \delta_{14}-2  \delta_{24}) \sin \delta_{13} s^2_{23} \right) s^2_{24} \right) c^2_{13} + 2 \cos \delta_{13} c_{12} c_{34} s_{2(13)} s_{23} \left( c_{12} s_{14} \left( c_{13} c_{24} s_{23}  \right. \right.\\
 & 
 \left(\sin \delta_{14} c^2_{24} +\sin ( \delta_{14}-2  \delta_{24}) s^2_{24} \right)- s_{13} s_{14} s_{24} \left(\cos ( \delta_{13}- \delta_{14}+ \delta_{24}) \sin \delta_{14} c^2_{24} +\sin ( \delta_{13}- \delta_{24}) 
   \right. \\
& \left. \left.\left. 
 s^2_{24} \right)\right)-c^2_{24} \sin \delta_{24} s_{12} s_{2(23)} s_{24} \right)  + s_{23} \left( c_{12} \sin \delta_{13} s_{13} \left(4 c_{12} c_{34}s_{13} s_{24} \left(\left( \sin \delta_{14} \sin ( \delta_{13}-  \delta_{14}+ \right. \right. \right. \right.\\
& \left. \left. \left. 
 \delta_{24}) s^2_{14} -\sin \delta_{13} \sin \delta_{24} s^2_{23}\right) c^2_{24} +\cos ( \delta_{13}- \delta_{24}) s^2_{14} s^2_{24} \right)+\left(( c_{2(14)} -3) c_{2(24)} -2 c^2_{14} \right) s_{12} s_{34} \right) \\
& \left. 
 - 2 c^2_{23} s_{2(24)} c_{24}  c_{34} s^2_{12} \sin \delta_{24} \right) c_{13} - 2 \cos ^2 \delta_{13} c^2_{12} s_{2(24)} c_{34} s^2_{13} s^2_{23}   ( c_{13} c_{24} \sin  \delta_{24} s_{23} +\sin ( \delta_{13}- \delta_{14})  \\
& 
 s_{13} s_{14} s_{24})+ c_{34} s_{14} s_{24} \left( s_{24} \left(-4 c_{24}  s_{13} \sin ( \delta_{13}- \delta_{14}) \left( c^2_{23} s^2_{12} + c^2_{12} \sin ^2 \delta_{13} s^2_{13} s^2_{23} \right)-\cos ( \delta_{14}- \right. \right. \\
& \left. 
 2  \delta_{24}) c_{23}  \sin ( \delta_{13}- \delta_{14}+ \delta_{24}) s_{2(12)} s_{2(13)} s_{14} s_{24} \right)+\cos ( \delta_{13}- \delta_{14}+ \delta_{24}) c_{23} s_{2(12)} s_{2(13)} s_{14} \left(\cos ( \delta_{14} \right. \\ 
& \left. \left. 
 -\delta_{24}) \sin \delta_{24} s^2_{24} - c^2_{24} \sin \delta_{14} \right)\right)+\frac{1}{2} s_{14} \left( s_{2(12)} \left(2 \sin (2  \delta_{13}- \delta_{14}+ \delta_{24}) s^2_{13} s^2_{23} +(c_{2(23)} - c_{2(13)} ) \right. \right. \\
& \left. \left. \left. \left.
 \sin ( \delta_{14}- \delta_{24})\right)-2 c_{2(12)} \sin ( \delta_{13}- \delta_{14}+ \delta_{24}) s_{13} s_{2(23)} \right) s_{2(24)} s_{34}\right]\right\} .
\end{align*}

%%%%%%%%%%bibliography%%%%%%%%%%%%%
\bibliography{sterileNu}

\begin{thebibliography}{10}
\providecommand{\url}[1]{\texttt{#1}}
\providecommand{\urlprefix}{URL }
\providecommand{\eprint}[2][]{\url{#2}}

\bibitem{Aguilar:2001ty}
A.~Aguilar-Arevalo et~al. (LSND), \emph{{Evidence for neutrino oscillations
  from the observation of anti-neutrino(electron) appearance in a
  anti-neutrino(muon) beam}},
  \MYhref[journalLinks]{http://dx.doi.org/10.1103/PhysRevD.64.112007}{Phys.
  Rev.
  }\MYhref[journalLinks]{http://dx.doi.org/10.1103/PhysRevD.64.112007}{\textbf{D64}
  (2001) 112007},
  \MYhref[eprintLinks]{http://arxiv.org/abs/hep-ex/0104049}{{\ttfamily
  arXiv:hep-ex/0104049 [hep-ex]}}.

\bibitem{Aguilar-Arevalo:2018gpe}
A.~A. Aguilar-Arevalo et~al. (MiniBooNE), \emph{{Significant Excess of
  ElectronLike Events in the MiniBooNE Short-Baseline Neutrino Experiment}},
  \MYhref[journalLinks]{http://dx.doi.org/10.1103/PhysRevLett.121.221801}{Phys.
  Rev. Lett.
  }\MYhref[journalLinks]{http://dx.doi.org/10.1103/PhysRevLett.121.221801}{\textbf{121}
  (2018) 22 221801},
  \MYhref[eprintLinks]{http://arxiv.org/abs/1805.12028}{{\ttfamily
  arXiv:1805.12028 [hep-ex]}}.

\bibitem{Mention:2011rk}
G.~Mention et~al., \emph{{The Reactor Antineutrino Anomaly}},
  \MYhref[journalLinks]{http://dx.doi.org/10.1103/PhysRevD.83.073006}{Phys.
  Rev.
  }\MYhref[journalLinks]{http://dx.doi.org/10.1103/PhysRevD.83.073006}{\textbf{D83}
  (2011) 073006},
  \MYhref[eprintLinks]{http://arxiv.org/abs/1101.2755}{{\ttfamily
  arXiv:1101.2755 [hep-ex]}}.

\bibitem{Mueller:2011nm}
T.~A. Mueller et~al., \emph{{Improved Predictions of Reactor Antineutrino
  Spectra}},
  \MYhref[journalLinks]{http://dx.doi.org/10.1103/PhysRevC.83.054615}{Phys.
  Rev.
  }\MYhref[journalLinks]{http://dx.doi.org/10.1103/PhysRevC.83.054615}{\textbf{C83}
  (2011) 054615},
  \MYhref[eprintLinks]{http://arxiv.org/abs/1101.2663}{{\ttfamily
  arXiv:1101.2663 [hep-ex]}}.

\bibitem{Huber:2011wv}
P.~Huber, \emph{{On the determination of anti-neutrino spectra from nuclear
  reactors}},
  \MYhref[journalLinks]{http://dx.doi.org/10.1103/PhysRevC.85.029901,
  10.1103/PhysRevC.84.024617}{Phys. Rev.
  }\MYhref[journalLinks]{http://dx.doi.org/10.1103/PhysRevC.85.029901,
  10.1103/PhysRevC.84.024617}{\textbf{C84} (2011) 024617}, [Erratum: Phys.
  Rev.C85,029901(2012)],
  \MYhref[eprintLinks]{http://arxiv.org/abs/1106.0687}{{\ttfamily
  arXiv:1106.0687 [hep-ph]}}.

\bibitem{Hampel:1998xg}
W.~Hampel et~al. (GALLEX), \emph{{GALLEX solar neutrino observations: Results
  for GALLEX IV}},
  \MYhref[journalLinks]{http://dx.doi.org/10.1016/S0370-2693(98)01579-2}{Phys.
  Lett.
  }\MYhref[journalLinks]{http://dx.doi.org/10.1016/S0370-2693(98)01579-2}{\textbf{B447}
  (1999) 127--133}.

\bibitem{Kaether:2010ag}
F.~Kaether et~al., \emph{{Reanalysis of the GALLEX solar neutrino flux and
  source experiments}},
  \MYhref[journalLinks]{http://dx.doi.org/10.1016/j.physletb.2010.01.030}{Phys.
  Lett.
  }\MYhref[journalLinks]{http://dx.doi.org/10.1016/j.physletb.2010.01.030}{\textbf{B685}
  (2010) 47--54},
  \MYhref[eprintLinks]{http://arxiv.org/abs/1001.2731}{{\ttfamily
  arXiv:1001.2731 [hep-ex]}}.

\bibitem{Abdurashitov:1999zd}
J.~N. Abdurashitov et~al. (SAGE), \emph{{Measurement of the solar neutrino
  capture rate with gallium metal}},
  \MYhref[journalLinks]{http://dx.doi.org/10.1103/PhysRevC.60.055801}{Phys.
  Rev.
  }\MYhref[journalLinks]{http://dx.doi.org/10.1103/PhysRevC.60.055801}{\textbf{C60}
  (1999) 055801},
  \MYhref[eprintLinks]{http://arxiv.org/abs/astro-ph/9907113}{{\ttfamily
  arXiv:astro-ph/9907113 [astro-ph]}}.

\bibitem{Abdurashitov:2009tn}
J.~N. Abdurashitov et~al. (SAGE), \emph{{Measurement of the solar neutrino
  capture rate with gallium metal. III: Results for the 2002--2007 data-taking
  period}},
  \MYhref[journalLinks]{http://dx.doi.org/10.1103/PhysRevC.80.015807}{Phys.
  Rev.
  }\MYhref[journalLinks]{http://dx.doi.org/10.1103/PhysRevC.80.015807}{\textbf{C80}
  (2009) 015807},
  \MYhref[eprintLinks]{http://arxiv.org/abs/0901.2200}{{\ttfamily
  arXiv:0901.2200 [nucl-ex]}}.

\bibitem{Acciarri:2016crz}
R.~Acciarri et~al. (DUNE), \emph{{Long-Baseline Neutrino Facility (LBNF) and
  Deep Underground Neutrino Experiment (DUNE)}}  (2016),
  \MYhref[eprintLinks]{http://arxiv.org/abs/1601.05471}{{\ttfamily
  arXiv:1601.05471 [physics.ins-det]}}.

\bibitem{Acciarri:2015uup}
R.~Acciarri et~al. (DUNE), \emph{{Long-Baseline Neutrino Facility (LBNF) and
  Deep Underground Neutrino Experiment (DUNE)}}  (2015),
  \MYhref[eprintLinks]{http://arxiv.org/abs/1512.06148}{{\ttfamily
  arXiv:1512.06148 [physics.ins-det]}}.

\bibitem{Strait:2016mof}
J.~Strait et~al. (DUNE), \emph{{Long-Baseline Neutrino Facility (LBNF) and Deep
  Underground Neutrino Experiment (DUNE)}}  (2016),
  \MYhref[eprintLinks]{http://arxiv.org/abs/1601.05823}{{\ttfamily
  arXiv:1601.05823 [physics.ins-det]}}.

\bibitem{Acciarri:2016ooe}
R.~Acciarri et~al. (DUNE), \emph{{Long-Baseline Neutrino Facility (LBNF) and
  Deep Underground Neutrino Experiment (DUNE)}}  (2016),
  \MYhref[eprintLinks]{http://arxiv.org/abs/1601.02984}{{\ttfamily
  arXiv:1601.02984 [physics.ins-det]}}.

\bibitem{Ayres:2007tu}
D.~S. Ayres et~al. (NOvA),
  \MYhref[journalLinks]{http://dx.doi.org/10.2172/935497}{\emph{{The NOvA
  Technical Design Report}}
  }\MYhref[journalLinks]{http://dx.doi.org/10.2172/935497}{ (2007)}.

\bibitem{Abe:2015zbg}
K.~Abe et~al. (Hyper-Kamiokande Proto-Collaboration), \emph{{Physics potential
  of a long-baseline neutrino oscillation experiment using a J-PARC neutrino
  beam and Hyper-Kamiokande}},
  \MYhref[journalLinks]{http://dx.doi.org/10.1093/ptep/ptv061}{PTEP
  }\MYhref[journalLinks]{http://dx.doi.org/10.1093/ptep/ptv061}{\textbf{2015}
  (2015) 053C02},
  \MYhref[eprintLinks]{http://arxiv.org/abs/1502.05199}{{\ttfamily
  arXiv:1502.05199 [hep-ex]}}.

\bibitem{Liu:1998qp}
C.~Liu and J.-H. Song, \emph{{Four light neutrinos in singular seesaw mechanism
  with Abelian flavor symmetry}},
  \MYhref[journalLinks]{http://dx.doi.org/10.1103/PhysRevD.60.036002}{Phys.
  Rev.
  }\MYhref[journalLinks]{http://dx.doi.org/10.1103/PhysRevD.60.036002}{\textbf{D60}
  (1999) 036002},
  \MYhref[eprintLinks]{http://arxiv.org/abs/hep-ph/9812381}{{\ttfamily
  arXiv:hep-ph/9812381 [hep-ph]}}.

\bibitem{Barry:2011wb}
J.~Barry, W.~Rodejohann and H.~Zhang, \emph{{Light Sterile Neutrinos: Models
  and Phenomenology}},
  \MYhref[journalLinks]{http://dx.doi.org/10.1007/JHEP07(2011)091}{JHEP
  }\MYhref[journalLinks]{http://dx.doi.org/10.1007/JHEP07(2011)091}{\textbf{07}
  (2011) 091}, \MYhref[eprintLinks]{http://arxiv.org/abs/1105.3911}{{\ttfamily
  arXiv:1105.3911 [hep-ph]}}.

\bibitem{Kawai:2019uei}
S.~Kawai and N.~Okada, \emph{{eV-scale sterile neutrinos from an extra
  dimension}}  (2019),
  \MYhref[eprintLinks]{http://arxiv.org/abs/1910.02936}{{\ttfamily
  arXiv:1910.02936 [hep-ph]}}.

\bibitem{Berryman:2015nua}
J.~M. Berryman, A.~de~Gouvêa, K.~J. Kelly and A.~Kobach, \emph{{Sterile
  neutrino at the Deep Underground Neutrino Experiment}},
  \MYhref[journalLinks]{http://dx.doi.org/10.1103/PhysRevD.92.073012}{Phys.
  Rev.
  }\MYhref[journalLinks]{http://dx.doi.org/10.1103/PhysRevD.92.073012}{\textbf{D92}
  (2015) 7 073012},
  \MYhref[eprintLinks]{http://arxiv.org/abs/1507.03986}{{\ttfamily
  arXiv:1507.03986 [hep-ph]}}.

\bibitem{Gandhi:2015xza}
R.~Gandhi, B.~Kayser, M.~Masud and S.~Prakash, \emph{{The impact of sterile
  neutrinos on CP measurements at long baselines}},
  \MYhref[journalLinks]{http://dx.doi.org/10.1007/JHEP11(2015)039}{JHEP
  }\MYhref[journalLinks]{http://dx.doi.org/10.1007/JHEP11(2015)039}{\textbf{11}
  (2015) 039}, \MYhref[eprintLinks]{http://arxiv.org/abs/1508.06275}{{\ttfamily
  arXiv:1508.06275 [hep-ph]}}.

\bibitem{Agarwalla:2016mrc}
S.~K. Agarwalla, S.~S. Chatterjee, A.~Dasgupta and A.~Palazzo, \emph{{Discovery
  Potential of T2K and NOvA in the Presence of a Light Sterile Neutrino}},
  \MYhref[journalLinks]{http://dx.doi.org/10.1007/JHEP02(2016)111}{JHEP
  }\MYhref[journalLinks]{http://dx.doi.org/10.1007/JHEP02(2016)111}{\textbf{02}
  (2016) 111}, \MYhref[eprintLinks]{http://arxiv.org/abs/1601.05995}{{\ttfamily
  arXiv:1601.05995 [hep-ph]}}.

\bibitem{Agarwalla:2016xxa}
S.~K. Agarwalla, S.~S. Chatterjee and A.~Palazzo, \emph{{Physics Reach of DUNE
  with a Light Sterile Neutrino}},
  \MYhref[journalLinks]{http://dx.doi.org/10.1007/JHEP09(2016)016}{JHEP
  }\MYhref[journalLinks]{http://dx.doi.org/10.1007/JHEP09(2016)016}{\textbf{09}
  (2016) 016}, \MYhref[eprintLinks]{http://arxiv.org/abs/1603.03759}{{\ttfamily
  arXiv:1603.03759 [hep-ph]}}.

\bibitem{Dutta:2016glq}
D.~Dutta et~al., \emph{{Capabilities of long-baseline experiments in the
  presence of a sterile neutrino}},
  \MYhref[journalLinks]{http://dx.doi.org/10.1007/JHEP11(2016)122}{JHEP
  }\MYhref[journalLinks]{http://dx.doi.org/10.1007/JHEP11(2016)122}{\textbf{11}
  (2016) 122}, \MYhref[eprintLinks]{http://arxiv.org/abs/1607.02152}{{\ttfamily
  arXiv:1607.02152 [hep-ph]}}.

\bibitem{Kelly:2017kch}
K.~J. Kelly, \emph{{Searches for new physics at the Hyper-Kamiokande
  experiment}},
  \MYhref[journalLinks]{http://dx.doi.org/10.1103/PhysRevD.95.115009}{Phys.
  Rev.
  }\MYhref[journalLinks]{http://dx.doi.org/10.1103/PhysRevD.95.115009}{\textbf{D95}
  (2017) 11 115009},
  \MYhref[eprintLinks]{http://arxiv.org/abs/1703.00448}{{\ttfamily
  arXiv:1703.00448 [hep-ph]}}.

\bibitem{Choubey:2017cba}
S.~Choubey, D.~Dutta and D.~Pramanik, \emph{{Imprints of a light Sterile
  Neutrino at DUNE, T2HK and T2HKK}},
  \MYhref[journalLinks]{http://dx.doi.org/10.1103/PhysRevD.96.056026}{Phys.
  Rev.
  }\MYhref[journalLinks]{http://dx.doi.org/10.1103/PhysRevD.96.056026}{\textbf{D96}
  (2017) 5 056026},
  \MYhref[eprintLinks]{http://arxiv.org/abs/1704.07269}{{\ttfamily
  arXiv:1704.07269 [hep-ph]}}.

\bibitem{Coloma:2017ptb}
P.~Coloma, D.~V. Forero and S.~J. Parke, \emph{{DUNE Sensitivities to the
  Mixing between Sterile and Tau Neutrinos}},
  \MYhref[journalLinks]{http://dx.doi.org/10.1007/JHEP07(2018)079}{JHEP
  }\MYhref[journalLinks]{http://dx.doi.org/10.1007/JHEP07(2018)079}{\textbf{07}
  (2018) 079}, \MYhref[eprintLinks]{http://arxiv.org/abs/1707.05348}{{\ttfamily
  arXiv:1707.05348 [hep-ph]}}.

\bibitem{Choubey:2017ppj}
S.~Choubey, D.~Dutta and D.~Pramanik, \emph{{Measuring the Sterile Neutrino CP
  Phase at DUNE and T2HK}},
  \MYhref[journalLinks]{http://dx.doi.org/10.1140/epjc/s10052-018-5816-y}{Eur.
  Phys. J.
  }\MYhref[journalLinks]{http://dx.doi.org/10.1140/epjc/s10052-018-5816-y}{\textbf{C78}
  (2018) 4 339},
  \MYhref[eprintLinks]{http://arxiv.org/abs/1711.07464}{{\ttfamily
  arXiv:1711.07464 [hep-ph]}}.

\bibitem{Gupta:2018qsv}
S.~Gupta, Z.~M. Matthews, P.~Sharma and A.~G. Williams, \emph{{The Effect of a
  Light Sterile Neutrino at NO$\nu$A and DUNE}},
  \MYhref[journalLinks]{http://dx.doi.org/10.1103/PhysRevD.98.035042}{Phys.
  Rev.
  }\MYhref[journalLinks]{http://dx.doi.org/10.1103/PhysRevD.98.035042}{\textbf{D98}
  (2018) 3 035042},
  \MYhref[eprintLinks]{http://arxiv.org/abs/1804.03361}{{\ttfamily
  arXiv:1804.03361 [hep-ph]}}.

\bibitem{Choubey:2018kqq}
S.~Choubey, D.~Dutta and D.~Pramanik, \emph{{Exploring fake solutions in the
  sterile neutrino sector at long-baseline experiments}},
  \MYhref[journalLinks]{http://dx.doi.org/10.1140/epjc/s10052-019-7479-8}{Eur.
  Phys. J.
  }\MYhref[journalLinks]{http://dx.doi.org/10.1140/epjc/s10052-019-7479-8}{\textbf{C79}
  (2019) 11 968},
  \MYhref[eprintLinks]{http://arxiv.org/abs/1811.08684}{{\ttfamily
  arXiv:1811.08684 [hep-ph]}}.

\bibitem{Betti:2019ouf}
M.~G. Betti et~al. (PTOLEMY), \emph{{Neutrino physics with the PTOLEMY project:
  active neutrino properties and the light sterile case}},
  \MYhref[journalLinks]{http://dx.doi.org/10.1088/1475-7516/2019/07/047}{JCAP
  }\MYhref[journalLinks]{http://dx.doi.org/10.1088/1475-7516/2019/07/047}{\textbf{1907}
  (2019) 047}, \MYhref[eprintLinks]{http://arxiv.org/abs/1902.05508}{{\ttfamily
  arXiv:1902.05508 [astro-ph.CO]}}.

\bibitem{Lindner:2015iaa}
M.~Lindner, W.~Rodejohann and X.-J. Xu, \emph{{Sterile neutrinos in the light
  of IceCube}},
  \MYhref[journalLinks]{http://dx.doi.org/10.1007/JHEP01(2016)124}{JHEP
  }\MYhref[journalLinks]{http://dx.doi.org/10.1007/JHEP01(2016)124}{\textbf{01}
  (2016) 124}, \MYhref[eprintLinks]{http://arxiv.org/abs/1510.00666}{{\ttfamily
  arXiv:1510.00666 [hep-ph]}}.

\bibitem{Abe:2019fyx}
K.~Abe et~al. (T2K), \emph{{Search for light sterile neutrinos with the T2K far
  detector Super-Kamiokande at a baseline of 295 km}},
  \MYhref[journalLinks]{http://dx.doi.org/10.1103/PhysRevD.99.071103}{Phys.
  Rev.
  }\MYhref[journalLinks]{http://dx.doi.org/10.1103/PhysRevD.99.071103}{\textbf{D99}
  (2019) 7 071103},
  \MYhref[eprintLinks]{http://arxiv.org/abs/1902.06529}{{\ttfamily
  arXiv:1902.06529 [hep-ex]}}.

\bibitem{Kamo:2002sj}
Y.~Kamo et~al., \emph{{Analytical calculations of four neutrino oscillations in
  matter}},
  \MYhref[journalLinks]{http://dx.doi.org/10.1140/epjc/s2003-01138-0}{Eur.
  Phys. J.
  }\MYhref[journalLinks]{http://dx.doi.org/10.1140/epjc/s2003-01138-0}{\textbf{C28}
  (2003) 211--221},
  \MYhref[eprintLinks]{http://arxiv.org/abs/hep-ph/0209097}{{\ttfamily
  arXiv:hep-ph/0209097 [hep-ph]}}.

\bibitem{Li:2018ezt}
W.~Li, J.~Ling, F.~Xu and B.~Yue, \emph{{Matter Effect of Light Sterile
  Neutrino: An Exact Analytical Approach}},
  \MYhref[journalLinks]{http://dx.doi.org/10.1007/JHEP10(2018)021}{JHEP
  }\MYhref[journalLinks]{http://dx.doi.org/10.1007/JHEP10(2018)021}{\textbf{10}
  (2018) 021}, \MYhref[eprintLinks]{http://arxiv.org/abs/1808.03985}{{\ttfamily
  arXiv:1808.03985 [hep-ph]}}.

\bibitem{Parke:2019jyu}
S.~J. Parke and X.~Zhang, \emph{{Compact Perturbative Expressions for
  Oscillations with Sterile Neutrinos in Matter}}  (2019),
  \MYhref[eprintLinks]{http://arxiv.org/abs/1905.01356}{{\ttfamily
  arXiv:1905.01356 [hep-ph]}}.

\bibitem{Abazajian:2012ys}
K.~N. Abazajian et~al., \emph{{Light Sterile Neutrinos: A White Paper}}
  (2012), \MYhref[eprintLinks]{http://arxiv.org/abs/1204.5379}{{\ttfamily
  arXiv:1204.5379 [hep-ph]}}.

\bibitem{Boser:2019rta}
S.~Böser et~al., \emph{{Status of Light Sterile Neutrino Searches}}  (2019),
  \MYhref[eprintLinks]{http://arxiv.org/abs/1906.01739}{{\ttfamily
  arXiv:1906.01739 [hep-ex]}}.

\bibitem{Diaz:2019fwt}
A.~Diaz et~al., \emph{{Where Are We With Light Sterile Neutrinos?}}  (2019),
  \MYhref[eprintLinks]{http://arxiv.org/abs/1906.00045}{{\ttfamily
  arXiv:1906.00045 [hep-ex]}}.

\bibitem{Tanabashi:2018oca}
M.~Tanabashi et~al. (Particle Data Group), \emph{{Review of Particle Physics}},
  \MYhref[journalLinks]{http://dx.doi.org/10.1103/PhysRevD.98.030001}{Phys.
  Rev.
  }\MYhref[journalLinks]{http://dx.doi.org/10.1103/PhysRevD.98.030001}{\textbf{D98}
  (2018) 3 030001}.

\bibitem{Esteban:2018azc}
I.~Esteban et~al., \emph{{Global analysis of three-flavour neutrino
  oscillations: synergies and tensions in the determination of $\theta_{23},
  \delta_{CP}$, and the mass ordering}},
  \MYhref[journalLinks]{http://dx.doi.org/10.1007/JHEP01(2019)106}{JHEP
  }\MYhref[journalLinks]{http://dx.doi.org/10.1007/JHEP01(2019)106}{\textbf{01}
  (2019) 106}, \MYhref[eprintLinks]{http://arxiv.org/abs/1811.05487}{{\ttfamily
  arXiv:1811.05487 [hep-ph]}}.

\bibitem{Zhang:2006yq}
H.~Zhang, \emph{{Sum rules of four-neutrino mixing in matter}},
  \MYhref[journalLinks]{http://dx.doi.org/10.1142/S0217732307022244}{Mod. Phys.
  Lett.
  }\MYhref[journalLinks]{http://dx.doi.org/10.1142/S0217732307022244}{\textbf{A22}
  (2007) 1341--1348},
  \MYhref[eprintLinks]{http://arxiv.org/abs/hep-ph/0606040}{{\ttfamily
  arXiv:hep-ph/0606040 [hep-ph]}}.

\bibitem{Jarlskog:1985ht}
C.~Jarlskog, \emph{{Commutator of the Quark Mass Matrices in the Standard
  Electroweak Model and a Measure of Maximal CP Violation}},
  \MYhref[journalLinks]{http://dx.doi.org/10.1103/PhysRevLett.55.1039}{Phys.
  Rev. Lett.
  }\MYhref[journalLinks]{http://dx.doi.org/10.1103/PhysRevLett.55.1039}{\textbf{55}
  (1985) 1039}.

\bibitem{Wu:1985ea}
D.-d. Wu, \emph{{The Rephasing Invariants and CP}},
  \MYhref[journalLinks]{http://dx.doi.org/10.1103/PhysRevD.33.860}{Phys. Rev.
  }\MYhref[journalLinks]{http://dx.doi.org/10.1103/PhysRevD.33.860}{\textbf{D33}
  (1986) 860}.

\bibitem{Guo:2001yt}
W.-l. Guo and Z.-z. Xing, \emph{{Rephasing invariants of CP and T violation in
  the four neutrino mixing models}},
  \MYhref[journalLinks]{http://dx.doi.org/10.1103/PhysRevD.65.073020}{Phys.
  Rev.
  }\MYhref[journalLinks]{http://dx.doi.org/10.1103/PhysRevD.65.073020}{\textbf{D65}
  (2002) 073020},
  \MYhref[eprintLinks]{http://arxiv.org/abs/hep-ph/0112121}{{\ttfamily
  arXiv:hep-ph/0112121 [hep-ph]}}.

\bibitem{Barger:1999hi}
V.~D. Barger, Y.-B. Dai, K.~Whisnant and B.-L. Young, \emph{{Neutrino mixing,
  CP and T violation, and textures in four neutrino models}},
  \MYhref[journalLinks]{http://dx.doi.org/10.1103/PhysRevD.59.113010}{Phys.
  Rev.
  }\MYhref[journalLinks]{http://dx.doi.org/10.1103/PhysRevD.59.113010}{\textbf{D59}
  (1999) 113010},
  \MYhref[eprintLinks]{http://arxiv.org/abs/hep-ph/9901388}{{\ttfamily
  arXiv:hep-ph/9901388 [hep-ph]}}.

\bibitem{Kalliomaki:1999ii}
A.~Kalliomaki, J.~Maalampi and M.~Tanimoto, \emph{{Search for CP violation at a
  neutrino factory in a four neutrino model}},
  \MYhref[journalLinks]{http://dx.doi.org/10.1016/S0370-2693(99)01252-6}{Phys.
  Lett.
  }\MYhref[journalLinks]{http://dx.doi.org/10.1016/S0370-2693(99)01252-6}{\textbf{B469}
  (1999) 179--187},
  \MYhref[eprintLinks]{http://arxiv.org/abs/hep-ph/9909301}{{\ttfamily
  arXiv:hep-ph/9909301 [hep-ph]}}.

\bibitem{Donini:1999he}
A.~Donini, M.~B. Gavela, P.~Hernandez and S.~Rigolin, \emph{{Four species
  neutrino oscillations at neutrino factory: Sensitivity and CP violation}},
  \MYhref[journalLinks]{http://dx.doi.org/10.1016/S0168-9002(00)00373-9}{Nucl.
  Instrum. Meth.
  }\MYhref[journalLinks]{http://dx.doi.org/10.1016/S0168-9002(00)00373-9}{\textbf{A451}
  (2000) 58--68},
  \MYhref[eprintLinks]{http://arxiv.org/abs/hep-ph/9910516}{{\ttfamily
  arXiv:hep-ph/9910516 [hep-ph]}}.

\bibitem{Xing:2001bg}
Z.-z. Xing, \emph{{Sum rules of neutrino masses and CP violation in the four
  neutrino mixing scheme}},
  \MYhref[journalLinks]{http://dx.doi.org/10.1103/PhysRevD.64.033005}{Phys.
  Rev.
  }\MYhref[journalLinks]{http://dx.doi.org/10.1103/PhysRevD.64.033005}{\textbf{D64}
  (2001) 033005},
  \MYhref[eprintLinks]{http://arxiv.org/abs/hep-ph/0102021}{{\ttfamily
  arXiv:hep-ph/0102021 [hep-ph]}}.

\end{thebibliography}
%%%%%%%%%%%%%%%%%%%%%%%%%%%%%%%%%%%

%%%%%%%%%%%%%%%%%%%%%%%
\end{document}